\documentclass[12pt]{spieman}  
\usepackage{amsmath,amsfonts,amssymb}
\usepackage{graphicx}
\usepackage{setspace}
\usepackage{tocloft}
\usepackage{siunitx}
\usepackage{threeparttable}
\usepackage{etoolbox}
\usepackage{lineno}
\appto\TPTnoteSettings{\footnotesize}
\usepackage{multirow}
\usepackage[table,xcdraw]{xcolor}

\usepackage{soul}

\title{Development of 13 \boldmath\textbf{}$\mu m$ Cutoff HgCdTe Detector Arrays for Astronomy}

\author[a,*]{Mario S. Cabrera}
\author[a]{Craig W. McMurtry}
\author[a]{Meghan L. Dorn}
\author[a]{William J. Forrest}
\author[a]{Judith L. Pipher}
\author[b]{Donald Lee}
\affil[a]{University of Rochester, Department of Physics and Astronomy, 500 Wilson Blvd., Rochester, NY, USA, 14627-0171}
\affil[b]{Teledyne Imaging Systems, 5212 Verdugo Way, Camarillo, CA, USA, 93012}

\cftpagenumbersoff{figure}
\cftpagenumbersoff{table} 
\begin{document} 
\maketitle

\begin{abstract}
Building on the successful development of the 10 $\mu m$ HgCdTe detector arrays for the proposed NEOCam mission, the University of Rochester Infrared Detector team and Teledyne Imaging Systems are working together to extend the cutoff wavelength of HgCdTe detector arrays initially to 13 $\mu m$, with the ultimate goal of developing 15 $\mu m$ HgCdTe detector arrays for space and ground-based astronomy. The advantage of HgCdTe detector arrays is that they can operate at higher temperatures than the currently used arsenic doped silicon detector arrays at the longer wavelengths. Our infrared detector team at the University of Rochester has received and tested four 13 $\mu m$ detector arrays from Teledyne Imaging Systems with three different pixel designs, two of which are meant to reduce quantum tunneling dark current. The pixel design of one of these arrays has mitigated the effects of quantum tunneling dark currents for which we have been able to achieve, at a temperature of 28 K and applied bias of 350 mV, a well depth of at least 75 $ke^-$ for 90\% of the pixels with a median dark current of 1.8 $e^-/sec$. These arrays have demonstrated encouraging results as we move forward to extending the cutoff wavelength to 15 $\mu m$.
\end{abstract}


{\noindent \footnotesize\textbf{*}Mario S. Cabrera,  \linkable{mcabrer2@ur.rochester.edu} }

\begin{spacing}{2}

\section{Introduction}
\label{sect:intro}

Cryogens or cryo-coolers used to cool long wave infrared (LWIR) detector arrays on space missions take up valuable space and weight, and the limited volume of cryogens limits the lifetime of the LWIR detector array sensitivity. This project addresses these limitations, since the current $\sim$13 $\mu m$ (LW13), and eventual 15 $\mu m$ (LW15) wavelength cutoff HgCdTe arrays, can operate at temperatures that can be attained through passive cooling in space.

Instruments in past space missions that used LWIR detector arrays (wavelength cutoff above $\sim$5$\mu m$) required cooling to very low temperatures with on-board cryogens. For example, the Spitzer Space Telescope's Si:As impurity band conduction (IBC) LWIR arrays with a cutoff wavelength of $\sim$28 $\mu m$ used in all three instruments (IRAC, MIPC, IRS)\cite{Werner04,Gehrz07} were operated at 6-8 K, and WISE's similar arrays (centered at 12 and 22 $\mu m$)\cite{Mainzer08} were cooled to 7.8 K. After cryogens ran out, due to good thermal design the focal plane of Spitzer Space Telescope warmed up and equilibrated to $\sim$ 28 K. The mid-wave IR InSb cameras continue to function at comparable sensitivity to that during the cryogenic phase, while the Si:As IBC LWIR large format detector arrays ceased to function due to high dark currents. Similarly, only the WISE mid-IR HgCdTe arrays continued to function once cryogen was depleted.

\begin{figure}[t]
\begin{center}
\begin{tabular}{c}
\includegraphics[scale=0.4]{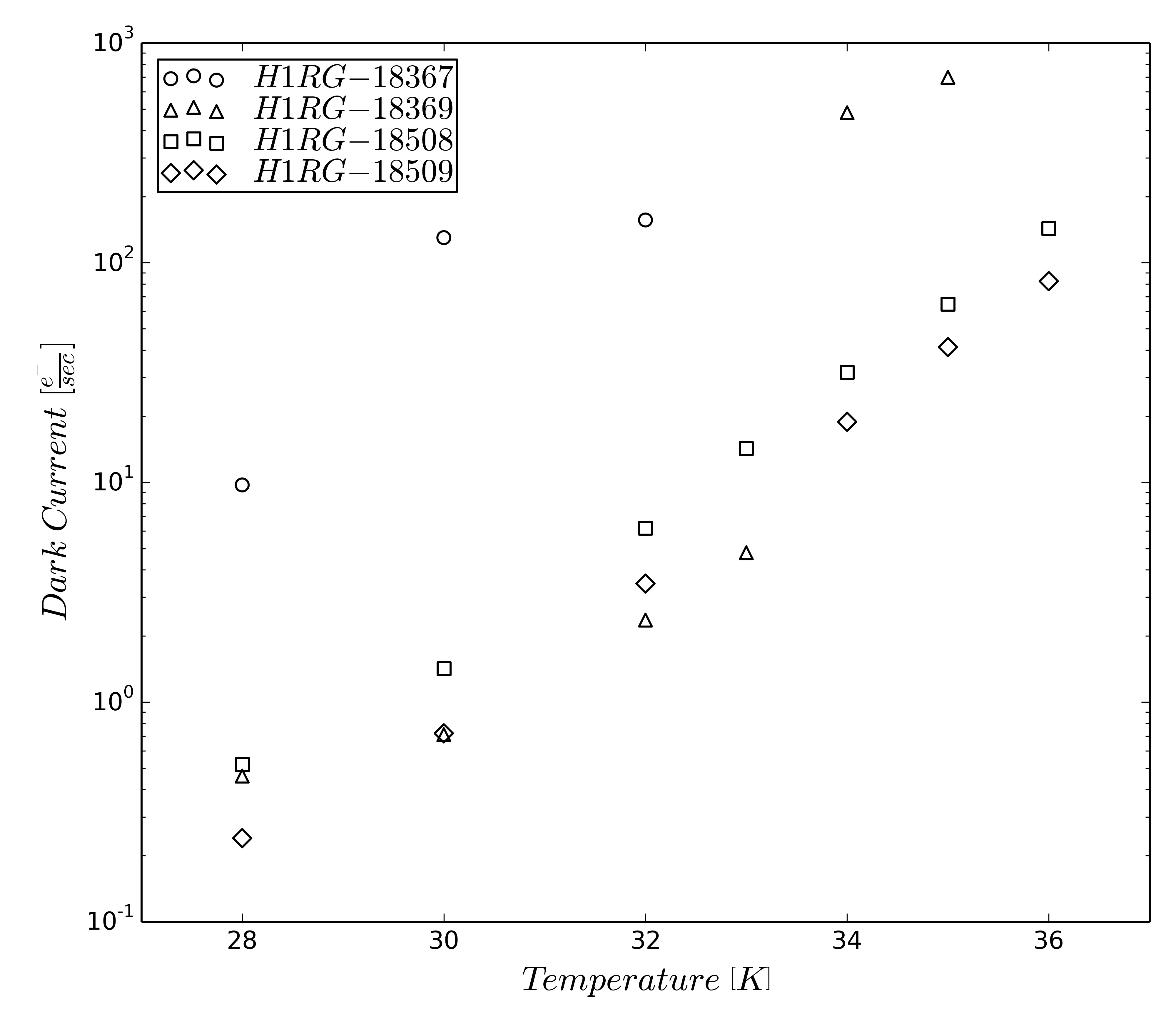}
\end{tabular}
\end{center}
\caption 
{ \label{fig:Ivstemp_150mV}
Median dark current \textit{vs.} temperature with 150 mV of applied bias for all four $\sim$13 $\mu m$ cutoff wavelength arrays presented here. The dark current measurements for H1RG-18367 at all temperatures, and H1RG-18369 at 34 and 35 K were affected by a glow from the multiplexer, see Sect. \ref{sect:mux glow} and \ref{sect:IvsWell}.} 
\end{figure}

Figure \ref{fig:Ivstemp_150mV} shows the median dark current \textit{vs.} temperature for all four $\sim$13 $\mu m$ cutoff wavelength HgCdTe arrays presented here, where three of the arrays show a median dark current $<$ 1 $e^-/sec$ at a temperature of 28 K and an applied bias of 150 mV.

Although the arrays presented here were developed for space missions, this technology can be important for ground based observatories. Instruments developed to explore the N-band that use 13 $\mu m$ arrays HgCdTe arrays, would be able to operate with the use of a cryo-cooler, eliminating the need for cryogens. We have initiated a study of this with colleagues, where our colleagues are experimenting with one of the LW13 arrays (H1RG-18508) to see if a fast readout (full-frame time of 12.5 msec) brings the N-band signal on scale, enabling astronomical observations with the array.


\subsection{Mercury Cadmium Telluride (HgCdTe)}
\label{sect:mercad}
Our group and Teledyne Imaging Systems (TIS) are working together to develop new 15 $\mu m$ LWIR detector arrays that can operate at relatively elevated focal plane temperatures for low background astronomy. TIS manufactures and develops these LWIR detector arrays with substantial feedback from our group. The testing and characterization of these devices is carried out at the University of Rochester (UR). The alloy used for these detectors, Hg$_{1-x}$Cd$_x$Te, has the $x$ composition parameter (molar concentration of Cd) varied to reach different cutoff wavelengths. The band gap energy (in eV) of this semiconductor at a temperature in degrees Kelvin is given by \cite{Hansen83}
\begin{equation}
\label{eq:E_g}
E_g(x,T) = -0.302 + 1.93x - 0.81x^2 + 0.832x^3 + 5.35\times10^{-4}T(1 - 2x)\, ,
\end{equation}
while the relationship between the band gap energy and the cut-off wavelength  is given by
\begin{equation}
\label{eq:E_g2}
E_g = \frac{hc}{\lambda_c}\, ,
\end{equation}
where $h$ is Planck's constant and $c$ is the speed of light.

\subsection{Prior Development / NEOCam}
\label{sect:NEOCam}
The UR infrared detector team has been working on the development and improvement of low background devices with a cut-off wavelength longer than about 5.4 $\mu m$ since 1992 \cite{Wu97,Bailey98,Bacon04,Bacon06,Bacon10}. By 2003, these early large format arrays were 512$\times$512 pixels and had cut-off wavelengths of 9.3 and 10.3 $\mu m$ (LW10 arrays), and showed low dark current, but could not support a large reverse bias which limited the well depth of these devices. Even with these limitations, the devices showed promise as they were both operable at a temperature of 30 K \cite{Carmody03,Bailey98}.

The Near Earth Object Camera (NEOCam) is a JPL proposed mission whose goal is to find and characterize near-Earth objects (asteroids and comets), some of which have Earth-crossing orbits and could be potentially hazardous. The mission will use the LW10 detector arrays discussed above to allow the telescope to observe from 6-10 $\mu m$. In addition to the instrumentation involving the LW10 arrays, NEOCam will also have a camera with lower cut-off wavelength arrays (5 $\mu m$) already developed for WISE and the James Webb Space Telescope \cite{Smith09_JWST}.

To meet the requirements of the NEOCam project, TIS successfully increased the array format from 512$\times$512 pixels to 2048$\times$2048 pixels. The arrays have cut-off wavelengths of $\sim$ 10 $\mu m$, low read noise, high quantum efficiency (QE), and have operabilities (well depth of more than 44,000 $e^-$ and dark current less than 200 $e^-/s$ ) $>$ 90\% up to temperatures of 42 K, and have been proton irradiated and demonstrated the ability to withstand the cosmic ray radiation that they are expected to receive in space \cite{Dorn16,McMurtry16,Dorn18}. These devices have reached NASA Technical Readiness Level-6.

\subsection{LW13 Detector Arrays}
\label{sect:LW13 summary}

In order to reach the present 15 $\mu m$ goal (composition parameter $x$ $\sim$ 0.209, at a temperature of 30 K), we are completing an intermediate step by developing arrays with a cutoff wavelength of 13 $\mu m$ ($x$ $\sim$ 0.216, at a temperature of 30 K) to identify any problems that would prevent us from reaching the 15 $\mu m$ goal. As the compound becomes softer (which is a consequence of increasing the mercury fraction with increased wavelength), there is an increased likelihood of defects/dislocations \cite{Carmody03} that would contribute to trap-assisted tunneling currents, as well as an increase in direct band-to-band tunneling due to the smaller band gap.

We have received from TIS four 1024$\times$1024 pixel arrays  bonded to Hawaii-1RG multiplexers\cite{Montroy02,Loose02,Loose07} with a pixel pitch of 18 $\mu m$ and a wavelength cutoff of $\sim$13 $\mu m$ for the first phase of this project. Two of the four arrays, H1RG-18367 and H1RG-18508, were grown and processed in the same manner as the LW10 arrays for the proposed NEOCam mission, but extended to the desired 13 $\mu m$ cutoff wavelength. Both H1RG-18369 and H1RG-18509 use TIS proprietary experimental structures designed to reduce tunneling dark currents, designated as design 1 and design 2 respectively.

\begin{table}
\centering
\caption{Cut-off wavelength and QE for all four LW13 arrays. QE values are expected to increase if arrays had anti-reflective coating.}
\label{tab:QE and cutoff}
\begin{tabular}{|c|c|c|c|c|}
\hline
\begin{tabular}[c]{@{}c@{}}Detector\\ H1RG-\end{tabular} & \begin{tabular}[c]{@{}c@{}}Wafer\\ 2-\end{tabular} & Lot-Split                                                          & \begin{tabular}[c]{@{}c@{}}Cut-off\\ Wavelength\\ ($\mu m$)\end{tabular} & \begin{tabular}[c]{@{}c@{}}QE\\ (6-10 $\mu m$) \end{tabular} \\ \hline
18367 & 3757 & Standard & 12.8 & 74\% \\ \hline
18508 & 3755 & Standard & 12.7 & 73\% \\ \hline
18369 & 3763 & Design 1 & 12.4 & 72\% \\ \hline
18509 & 3759 & Design 2 & 12.6 & 73\% \\ \hline
\end{tabular}
\end{table}

TIS provided measurements of the QE before anti-reflection coating and cutoff wavelength from mini-arrays on process evaluation chips (PECs) manufactured with the arrays, at a temperature of 30 K (Table \ref{tab:QE and cutoff}). The QE as a function of wavelength was measured at UR using a circular variable filter, and we only quote the PEC QE measurements since we found a reasonable agreement with TIS measurements.

\subsubsection{Operability}
\label{sect:operability}

Operable pixels should have low initial dark current with sufficient well depth, high QE, and low read noise. The performance of the LW13 arrays has been shown to be limited by dark current and well depths during characterization.

These arrays have shown excellent QE ($>$ 70\% before anti-reflection coating) and low correlated double sample (CDS) read noise ($\sim$ 23 $e^-$ shown for two arrays in Section \ref{sect:read noise}) both at 30 K. We will show that high dark currents and/or low well depths limit the performance of these arrays. For this developmental project, we have focused our efforts in the reduction of high dark currents since their effects are expected to worsen when the wavelength is increased to the final goal of 15 $\mu m$. Therefore, pixels with low dark current and sufficient well depth will be considered operable.

We adopt the same dark current operablity requirements to that of the LW10 devices developed for the NEOCam mission \cite{Dorn18}. Our imposed operability requirements include dark currents $<$ 200 $e^-/s$, and well depth of $\sim$ 40 $ke^-$ for an applied bias of 150 mV (larger well depth requirements are used for larger applied biases). The dark current requirement for NEOCam was established such that the mission's 6-10 $\mu m$ channel will be background-limited by the thermal emission from the zodiacal dust cloud.

The operability requirements adopted for NEOCam are also specific to the filters, telescope size and throughput. Any future mission that intends to use these LW13 devices likely would have a modified set of requirements.

Though operability requirements depend on specific applications of these devices, the dark current and well depth requirements for operable pixels in the LW13 arrays are used as a benchmark to compare the performance of the different arrays at different temperatures and applied bias, and to determine the best pixel design that will be pursued when increasing the cutoff wavelength to 15 $\mu m$.

\subsubsection{Sources of Dark Current and their Implications}

Thermal (diffusion\cite{Reine81} and generation-recombination\cite{Sah57}) and tunneling\cite{Kinch81, Sze69, Kinch14} (band-to-band and trap-to-band) dark currents have been found to be the sources of dark current in these LW13 devices. Tunneling currents are the primary dark current mechanisms limiting the operability of the LW13 arrays at low temperatures and moderate high applied reverse bias required for use with low-power source-follower readout integrated circuit (ROIC) commonly used for astronomy focal plane arrays.

Both tunneling dark currents are strong functions of bias, and have large non-linear effects that can affect data calibration. In addition to the non-linear effects, the charge capacity of pixels with very large dark currents, due to trap-to-band tunneling, can be depleted leading to low well depths, where a subset of these pixels create a cross-hatching pattern in the operability map for all four arrays (see section \ref{sect:367_oper_discharge}).

The design of H1RG-18509 successfully lowered tunneling dark currents, where  the median dark current and well depth at a temperature of 28 K and reverse applied bias of 350 mV is 1.8 $e^-/s$ and 81 $ke^-$ respectively, while the median dark current for the other three arrays was $>$200 $e^-/s$. The effects of tunneling currents on calibration and operability for all four devices are shown in section \ref{sect:Characterization}.

The dark current of these devices has been modeled using the theory described in the following section. The results are shown in section \ref{sect:dark current results}, where we have shown that at biases $> \sim$200 mV and low temperatures, band-to-band tunneling is the dominant component of dark current and is shown to be fairly uniform since all pixels are affected equally by this dark current mechanism. 

Operating regimes in which thermal currents dominate would be ideal to avoid any non-linear dark current behavior as pixels debias due to integrating signal since both diffusion and generation-recombination dark currents do not vary significantly with a bias above 25 mV. These regimes will be further discussed in section \ref{sect:Discussion}.

\section{Dark Current Theory}
\label{sect:theory}

\subsection{Diffusion Current}
\label{sect:diffusion}
Diffusion dark current occurs when electrons in the valence band gain enough energy thermally to overcome the band gap and transition to the conduction band. This thermally induced current must occur within one diffusion length (much larger than the pixel width and depth in these arrays) from the depletion region, and is given by \cite{Reine81}
\begin{equation}
\label{eq:diffusion}
I_{dif} = A\frac{n_i^2 d}{N_d \tau_b }\left[exp\left(\frac{qV_{actual\ bias}}{k_b T}\right) - 1\right]\, ,
\end{equation}
where $A$ is the diode junction area, $n_i$ is the intrinsic carrier concentration, $d$ is the thickness of the n-type region, $N_d$ is the doping density, $\tau_b$ is the minority carrier (hole in the n-type region) lifetime, $k_b$ is Boltzmann’s constant, and T is the temperature. In reverse bias, the actual bias across the diode ($V_{actual\ bias}$) is negative and the exponential term is negligible for the applied biases that we typically apply ($>$ 25 mV of reverse bias). It can be readily seen that diffusion dark current does not change appreciably with varying bias, while the strong temperature dependence comes in the form of the intrinsic carrier concentration \cite{Hansen83}
\begin{equation}
\label{eq:n_i}
n_i = \left( 5.585 - 3.820x + 1.753 \times 10^{-3}T - 1.364 \times 10^{-3}xT \right) \times \left[ 10^{14} E_g^{3/4} T^{3/2} exp\left( - \frac{E_g}{2k_b T} \right) \right]\, ,
\end{equation}
where $x$ is the cadmium mole fraction.

\subsection{Generation-Recombination}
\label{sect:G-R}

The second source of dark current that increases with temperature is Generation-Recombination (G-R). Traps in the depletion region with energy levels between the valence band and the conduction band can facilitate the indirect transition of an electron to the conduction band, where electrons in traps would have to overcome a smaller energy gap. G-R current is given in Sah (1957) \cite{Sah57} by

\begin{equation}
\label{eq:g-r}
I_{G-R} = \frac{n_iW_DA}{\tau_{GR}}\left[\frac{sinh\left(\frac{-qV_{actual}}{2kT}\right)}{\frac{q\left(V_{bi}-V_{actual}\right)}{2kT}}\right]f(b),
\end{equation}

\begin{equation}
\label{eq:f}
f(b)=\int_{0}^{\infty}\frac{dz}{z^2+2bz+1},
\end{equation}

\begin{equation}
\label{eq:b}
b=exp\left[\frac{-qV_{actual}}{2kT}\right]\cosh\left(\frac{E_i-E_{t_{gr}}}{kT}\right).
\end{equation}
$E_{t_{gr}}$ is the trap energy level position with respect to the valence band which contributes the most to the G-R current, while $E_i$ is the intrinsic Fermi level ($E_g/2$). $\tau_{GR}$ is the depletion region lifetime for holes and electrons, and $W_D$ is the depletion region width given by \cite{Wu97}

\begin{equation}
\label{eq:depletion width}
W_D = \sqrt{\frac{2\epsilon \epsilon_0 \left( V_{bi} - V_{actual\ bias} \right)}{qN_d}},
\end{equation}
where $\epsilon_0$ is the permittivity of free space, $\epsilon$ is the relative permittivity of HgCdTe, and $V_{bi}$ is the built-in voltage. G-R current also has a very weak dependence on bias (as long as the back-bias is greater than $\sim$25 mV).

\subsection{Quantum Tunneling Currents}
\label{sect:tunneling}

Tunneling currents are produced by electrons tunneling from the valence to the conduction band directly (band-to-band), or by traps with energies between the valence and the conduction band (trap-to-band). There are two different models for band-to-band currents; the first assumes a triangular barrier, while the second one uses a parabolic barrier. The equation corresponding to the parabolic barrier is omitted here as our data best matches the behavior of a triangular barrier, and it is given by \cite{Kinch81,Sze69}
\begin{equation}
\label{eq:band-to-band}
I_{band-to-band} = -\frac{q^2AEV_{actual\ bias}}{4\pi^2 \hbar^2} \sqrt{\frac{2m_{eff}}{E_g}} exp\left( -\frac{4\sqrt{2m_{eff}}E_g^{3/2}}{3q\hbar E} \right)
\end{equation}
where $m_{eff}$ is the effective mass of the minority carrier and $E$ is the electric field across the depletion region given by \cite{Reine81}
\begin{equation}
\label{eq:E-field}
E = \sqrt{\frac{2N_d\left( E_g - qV_{actual\ bias} \right)}{\epsilon \epsilon_0}}.
\end{equation}

Trap-to-Band tunneling current for a triangular barrier is modeled by \cite{Kinch81,Kinch14}
\begin{equation}
\label{eq:trap-to-band}
I_{trap-to-band} = \frac{\pi^2 q m_{eff} A V_{actual\ bias} M^2 n_t}{h^3 \left( E_g - E_t\right)} exp \left( -\frac{4\sqrt{2 m_{eff}} \left(E_g-E_t\right)^{3/2}}{3qE\hbar} \right),
\end{equation}
where $M$ is the mass matrix, $E_t$ is the energy of the trap level with respect to the valence band, and $n_t$ is the trap density in the depletion region at $E_t$.

Data from pixels that appear to exhibit trap-to-band tunneling in the LW10\cite{Wu97,Bacon06,Bacon10} and the LW13 devices have also shown a soft breakdown in the I-V curves, prior to the onset of trap-to-band tunneling. Other authors have observed soft breakdowns in 4H-SiC\cite{Neudeck98,Neudeck99} and Silicon\cite{Ravi73} diodes caused by screw dislocations and stacking faults respectively, which became electrically active at a certain ``threshold voltage". Furthermore, Neudeck et al.\cite{Neudeck98}, Ravi et al.\cite{Ravi73}, and Benson et al.\cite{Benson10} (for HgCdTe on Si) showed that there is no correlation  between the degraded I-V characteristics and the trap density, because all traps associated with dislocations may not contribute to the soft breakdown due to the varying threshold voltage for dislocations within the same diode at which they become electrically active.

Though we cannot say with certainty what type of dislocation\cite{Carmody03} causes the soft breakdown in the LW13 arrays presented here, we believe a defect associated mechanism, likely different from that described by the authors above is responsible for this early soft breakdown and the onset of trap-to-band tunneling as traps become electrically active. Bacon (2006) \cite{Bacon06} parametrically fit the I-V curves by introducing a threshold voltage at which certain traps become active and contribute to trap-to-band tunneling as
\begin{equation}
\label{eq:n_t}
n_t = n_{t_i} + \frac{n_{t_d}}{1 + exp \left[\frac{\gamma q\left(V_a + V_{actual\ bias}\right)}{kT} \right]},
\end{equation}
where $n_{t_i}$ is an initial active trap density, $n_{t_d}$ is the trap density due to activated dislocations at a voltage $V_a$, and $\gamma$ is a parameter which dictates how sharply the current increases due to the soft breakdown before reaching the current expected from trap-to-band tunneling. A small modification to the Eq. 2.29 in Bacon (2006) was made where we multiply the sum of the activation voltage and the actual bias by $\gamma$. This change was made so that the fitted $\gamma$ parameter only affected the sharpness of the soft breakdown, and not the fitted activation voltage.

Although this is not a physical model since we do not know the trap identity, a more detailed study using this phenomenological fit can give us information about the traps such as the trap density and a distribution of trap energies. 

\begin{figure}
\begin{center}
\begin{tabular}{c}
\includegraphics[scale=0.45]{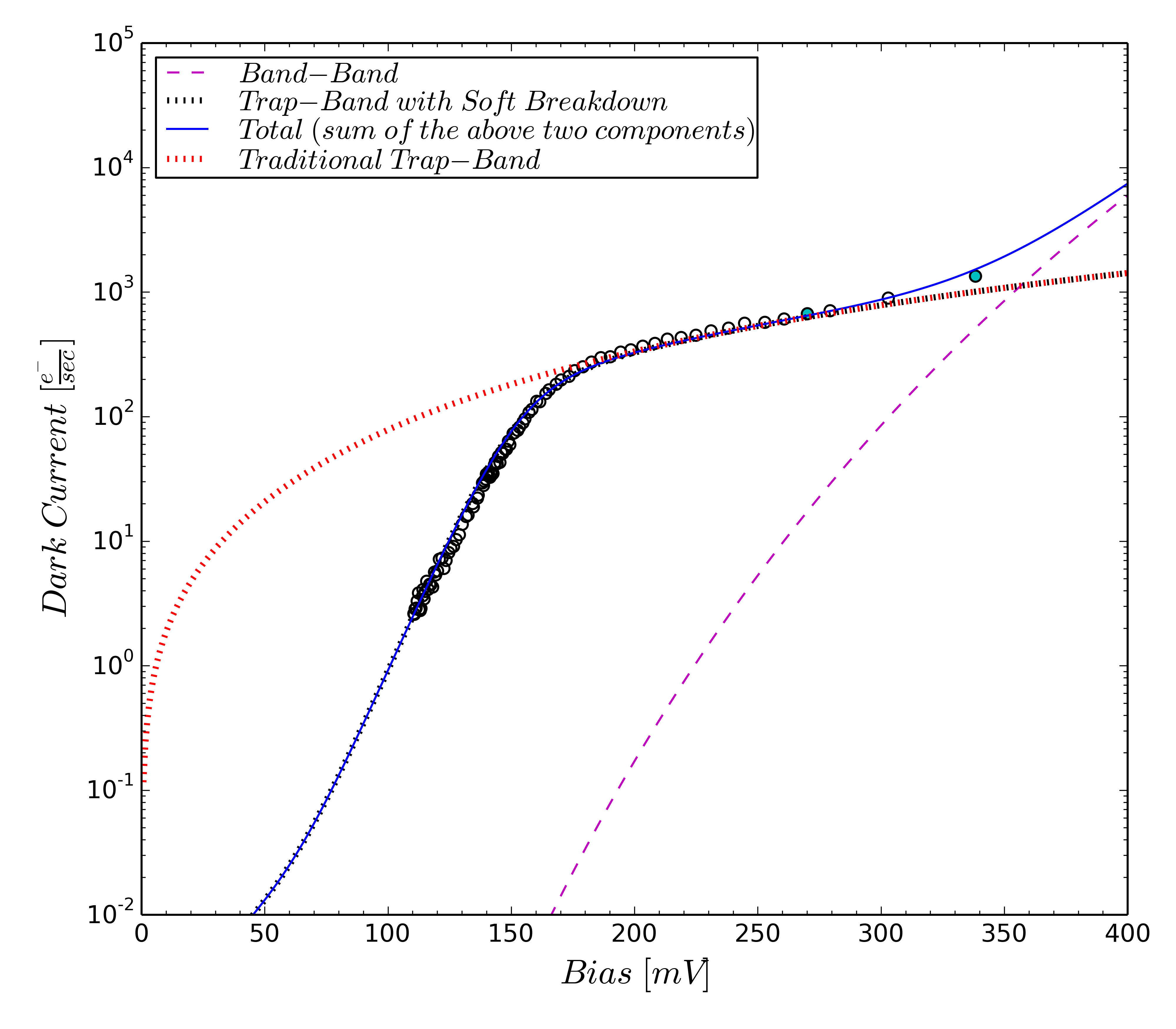}
\end{tabular}
\end{center}
\caption 
{ \label{fig:369_tradtraptoband}
Dark current \textit{vs.} bias at a temperature of 28 K for an inoperable pixel in H1RG-18369. Trap-to-band tunneling with a constant trap density and with variable trap density (due to activated traps) are plotted for comparison. The total curve corresponds to the sum of band-to-band and trap-to-band with soft breakdown curves.}
\end{figure}

Figure \ref{fig:369_tradtraptoband} shows the I-V data (see Section \ref{sect:I-V} for a description of how I-V data are obtained) of an inoperable pixel which can be modeled with trap-to-band tunneling following the soft breakdown. Below 170 mV of applied reverse bias, trap-to-band tunneling with the assumption of a constant trap density at all biases (labeled as ``Traditional Trap-Band'') does not fit the data, requiring the use of Eq. \ref{eq:n_t} which allows for the activation of traps. Similar behavior can be seen among inoperable pixels in all four LW13 arrays, where the data following the soft breakdown can be modeled by trap-to-band tunneling. 

More data at lower biases would be needed to more accurately characterize the threshold voltage of the traps which contribute to the trap-to-band tunneling, while data at larger biases is needed to properly characterize pixels which only appear to show the soft break down, and there is no indication if the behavior before or after the soft breakdown is consistent with trap-to-band tunneling or any other form of dark current.

\section{Data Acquisition}

UR uses a liquid helium dewar with multiple chambers to test the arrays. The innermost chamber that houses the array and filter wheel has an aluminum cylindrical shield coupled to a liquid helium reservoir, allowing temperatures down to 4 K and below (by reducing the pressure in the liquid helium reservoir) to be reached. The filter wheel has several narrow band filters, two circular variable filters (covering a wavelength range between 4-14.3 $\mu m$), and a dark blocking filter. Attached to the filter wheel housing is a 67.6 $\mu m$ diameter Lyot stop used to control the illumination of the array. The temperature inside the innermost chamber is regulated with a Lake Shore Cryotonics temperature controller, allowing the arrays presented here to be tested at different stable temperatures.

The inner chamber is surrounded by an aluminum cylindrical shield attached to a liquid nitrogen reservoir, shielding the liquid helium reservoir and inner chamber from the outer shell which is at room temperature. The outer shell has an anti-reflection coated ZnSe window in the line of sight of the array being tested.

The array controller used for data acquisition is based on an open source hardware design developed by the Observatory of the Carnegie Institute of Washington (OCIW). The version that we use is optimized for infrared arrays. It has programmable clocks and biases with four channels of amplification and analog to digital conversion for the output of the array. A detailed description of the array controller can be found in Moore et al. (2003) \cite{Moore03}.

\subsection{Sampling Modes}
\label{sampling methods}
The data presented here were obtained using either the Sample-Up-The-Ramp (SUTR) or Correlated Double Sample (CDS, or Fowler-1) sampling methods\cite{Fowler90,Garnett93,Rauscher07}, where the method used for specific data sets will be mentioned in the their description. The same reset mode is used for both sampling methods, where a reverse bias ($V_{bias}$) is applied to reset the array row by row (all columns in a row simultaneously). Immediately after the reset switch is turned off to allow the device to debias as signal and dark current are accumulated, a redistribution of charge (pedestal injection due to capacitive coupling to the reset FET) results in a change in the bias voltage across the diode to a value designated as $V_{actual}$.

Following the reset, the first sampled image is referred to as the pedestal. The delay time to the pedestal sample ranges from circa 5.8 milliseconds to 5.8 seconds, depending on the row and column for a specific pixel. For CDS images, the pedestal is subtracted from an image sampled some integration time later. The SUTR method is used to obtain many non-destructive samples to study the behavior of the array as pixels debias over time. In this sampling mode, following the reset and pedestal frame (used to normalize the rest of the samples to a zero point), many samples are taken at equal time intervals without resetting the array.

\section{Characterization}
\label{sect:Characterization}

To characterize the performance of the LW13 arrays we measured the dark current and well depth (amount of charge collected at the integrating node when saturation is reached) per pixel at temperatures ranging from 28 to 36 K with applied reverse bias of 150, 250, and 350 mV, as well as the read-noise per pixel at a temperature of 30 K for two of the arrays. The node capacitance and signal linearity have also been measured at a temperature of 30 K with applied reverse bias of 150, 250, and 350 mV to calibrate the arrays.

\subsection{Calibration}
\label{sect:calibration}

The first step in the calibration process is to measure the source-follower FET gain of signal in the multiplexer by turning on the reset switch, and varying the reset voltage while recording the output voltage. The source-follower FET gain was measured to be $\sim$0.9 for all four arrays, allowing us to convert between the output referred and input referred signal.

\begin{figure}
\begin{center}
\begin{tabular}{c}
\includegraphics[scale=0.5]{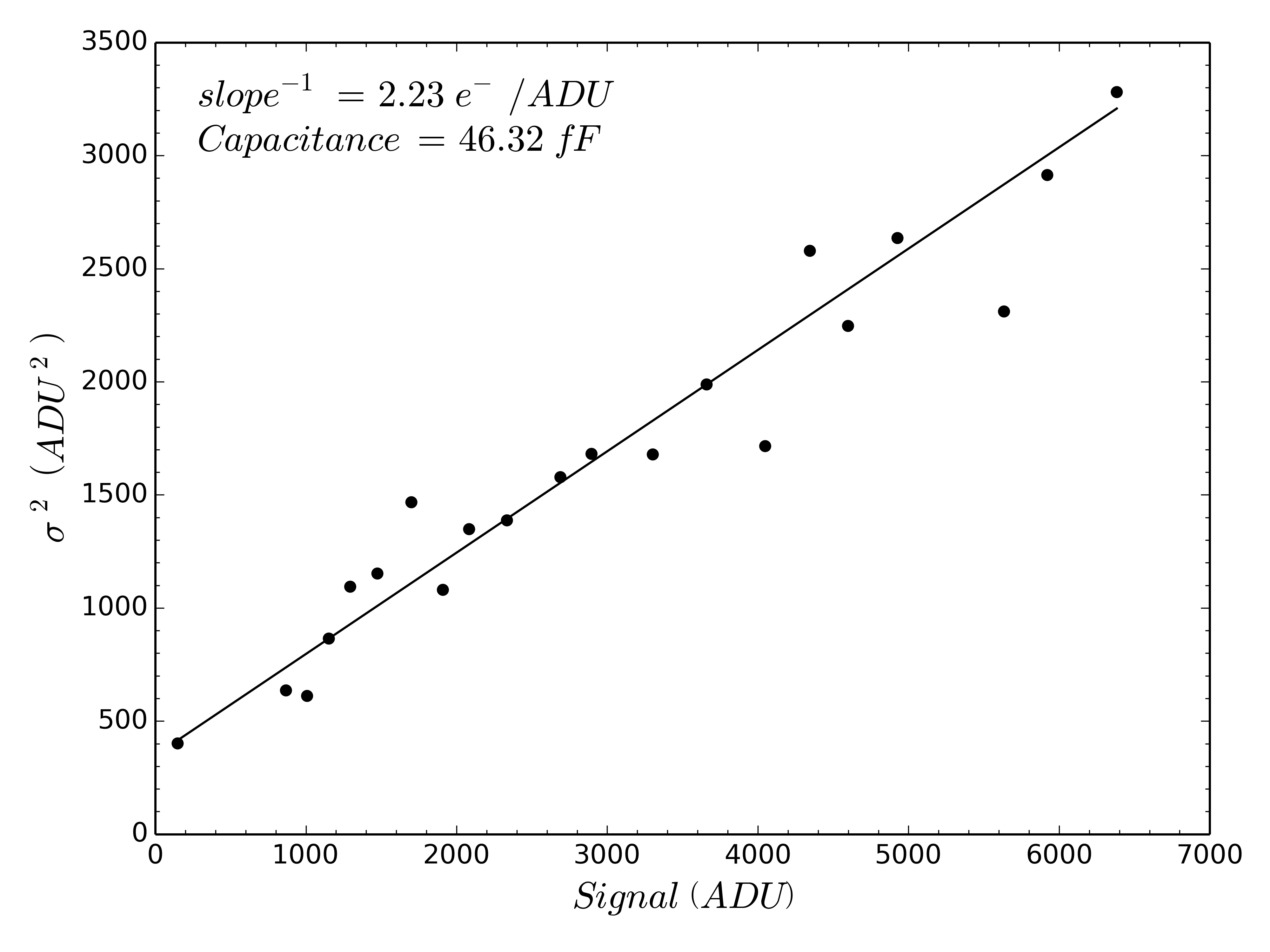}
\end{tabular}
\end{center}
\caption 
{ \label{fig:508_sig2vssig}
Noise squared \textit{vs.} signal plot for a pixel in H1RG-18508 at a temperature of 30K and 150 mV of applied bias, where the slope of the fitted line corresponds to the conversion factor between ADUs and $e^-$. The capacitance calculated here is not yet corrected for interpixel capacitance.} 
\end{figure}

\begin{figure}
\begin{center}
\begin{tabular}{c}
\includegraphics[scale=0.5]{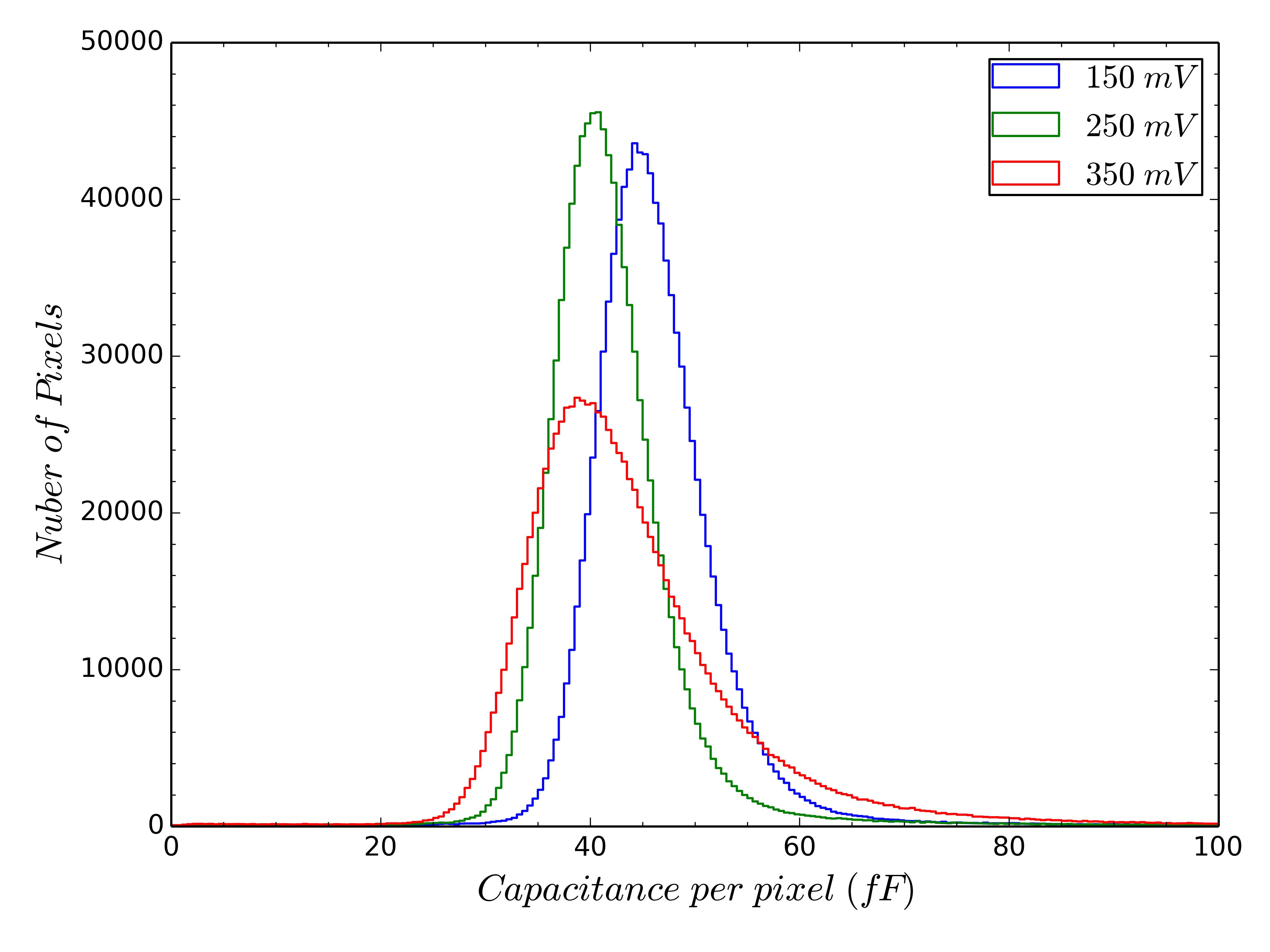}
\end{tabular}
\end{center}
\caption 
{ \label{fig:508_capperpix}
Capacitance per pixel distribution for H1RG-18508 at a temperature of 30 K and at all tested applied biases. These capacitances have not yet been corrected for interpixel capacitance. The spread in the capacitance distribution is due to noise. One standard deviation away from the mean for the 150 and 250 mV distributions is equal to 6.8 fF and 10.5 fF for the 350 mV distribution.} 
\end{figure}

Next, to convert from volts to electrons we measure the nodal capacitance. Our data are recorded in analog to digital units (ADU), which can be converted to volts by dividing the 5 V range of the 16 bit A/D converter by $2^{16}$ ADUs and the gain from our array controller electronics. The nodal capacitance per pixel is obtained using the noise squared ($\sigma^2$) \textit{vs.} signal method \cite{Mortara81} shown in Fig. \ref{fig:508_sig2vssig}. Sets of 100 CDS images, each set at varying fluence levels, were used to obtain the signal and the rms noise. Figure \ref{fig:508_capperpix} shows the nodal capacitance distribution per pixel for H1RG-18508 for three applied biases, where the median capacitance is then used to convert between the measured ADUs and the signal in electrons for the entire array.

The capacitance then has to be corrected for interpixel capacitance (IPC), determined through the nearest neighbor method. Pixels with a very high signal (due to very high dark currents) are used to determine the coupling parameter $\alpha$ between the high signal pixel and its nearest neighbors \cite{Moore04}. The nodal capacitance is then multiplied by a factor of $1-8\alpha$ to correct for IPC \cite{Moore04}. The IPC coupling factor $\alpha$ and corrected capacitance for all four arrays are shown in Table \ref{tab:IPC and Cap} for three applied biases.

\begin{table}
\centering
\caption{IPC coupling parameter $\alpha$ is given for each of the arrays, as well as the median IPC corrected capacitance in femtofarads for each of the applied biases.}
\label{tab:IPC and Cap}
\begin{tabular}{cc|c|c|c|}
\cline{3-5}
 &  & 150 mV & 250 mV & 350 mV \\ \hline
\multicolumn{1}{|c|}{\begin{tabular}[c]{@{}c@{}}Detector\\ H1RG-\end{tabular}} & $\alpha$ (\%) & \multicolumn{3}{c|}{\begin{tabular}[c]{@{}c@{}}Median Capacitance (fF)\\ Corrected for IPC\end{tabular}} \\ \hline
\multicolumn{1}{|c|}{18367} & 1.04 & 43 & 37 & 35 \\ \hline
\multicolumn{1}{|c|}{18508} & 1.03 & 42 & 38 & 38 \\ \hline
\multicolumn{1}{|c|}{18369} & 1.04 & 38 & 36 & 35 \\ \hline
\multicolumn{1}{|c|}{18509} & 1.12 & 39 & 37 & 34 \\ \hline
\end{tabular}
\end{table}

This method of converting the signal at the integrating node to electrons is only valid for low signal data which minimizes debiasing as in Fig. \ref{fig:508_sig2vssig}. For large signals, we need to account for the non-linear diode capacitance resulting from the debiasing of the diode, leading to the reduction of the junction depletion region as signal is collected \cite{Wu97_nonlinearity}. To measure the non-linearity of these devices, the signal collected from a constant illuminating flux is sampled in SUTR mode until saturation is reached. The non-linear collected signal rate is calculated by dividing the signal over the time it took to collect the signal. The signal is not corrected for dark current.

\begin{figure}[htb!]
\begin{center}
\begin{tabular}{c}
\includegraphics[scale=0.6]{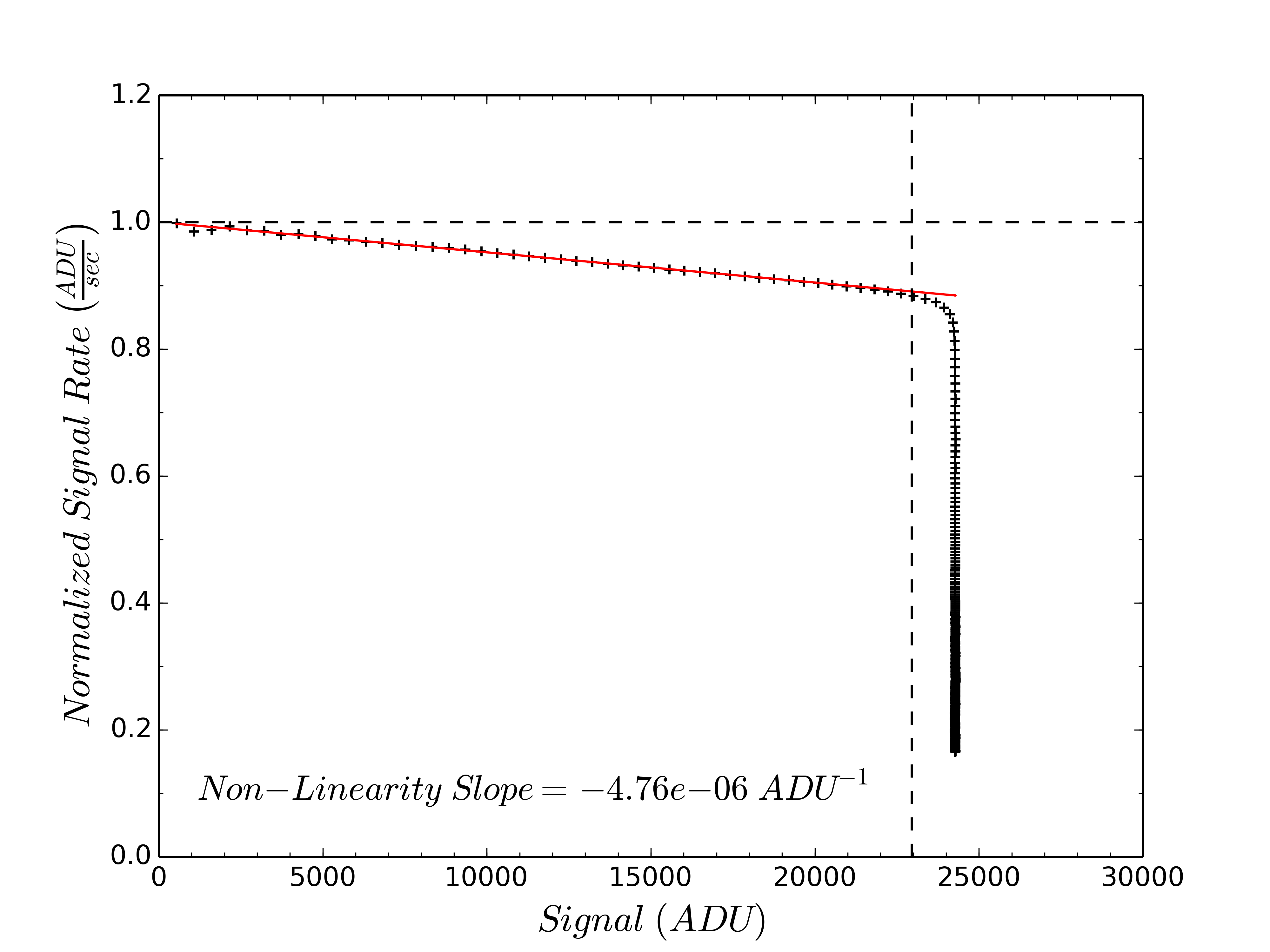}
\end{tabular}
\end{center}
\caption 
{ \label{fig:509_non-linearity_150mV}
Non-Linearity curve obtained from a 50X50 box average of pixels in H1RG-18509 with an applied bias of 150 mV at a temperature of 30 K. The collected signal rate was normalized to that corresponding to the lowest signal. The difference between a value of unity (horizontal dashed line) and the non-linearity curve is the fraction of the signal not collected due to the debiasing of the device. The vertical dashed line corresponds to the mean saturating signal (well depth) for the entire device, which occurs at actual zero bias. The saturation of the non-linearity curve occurs at a higher signal value due to the forward biasing of the detector by the signal flux used to saturate the array.} 
\end{figure}

Figure \ref{fig:509_non-linearity_150mV} shows the non-linearity curve for H1RG-18509 with an applied bias of 150 mV at 30 K, normalized to the first sampled signal. The normalized signal rate is also referred to as $C_0/C$, where $C_0$ is the nodal capacitance at zero collected signal and $C$ is the nodal capacitance. To correct for the non-linear capacitance, the slope of a fitted line to the non-linearity curve (between 20 and 80\% of the saturating signal) is used. Correcting for the non-linearity can be difficult when dark currents are on the order of the collected signal, which can occur at biases larger than 150 mV. A further step is required at high bias, where the extrapolated line fit is normalized to unity at zero signal.

\begin{figure}[htb!]
\begin{center}
\begin{tabular}{c}
\includegraphics[scale=0.6]{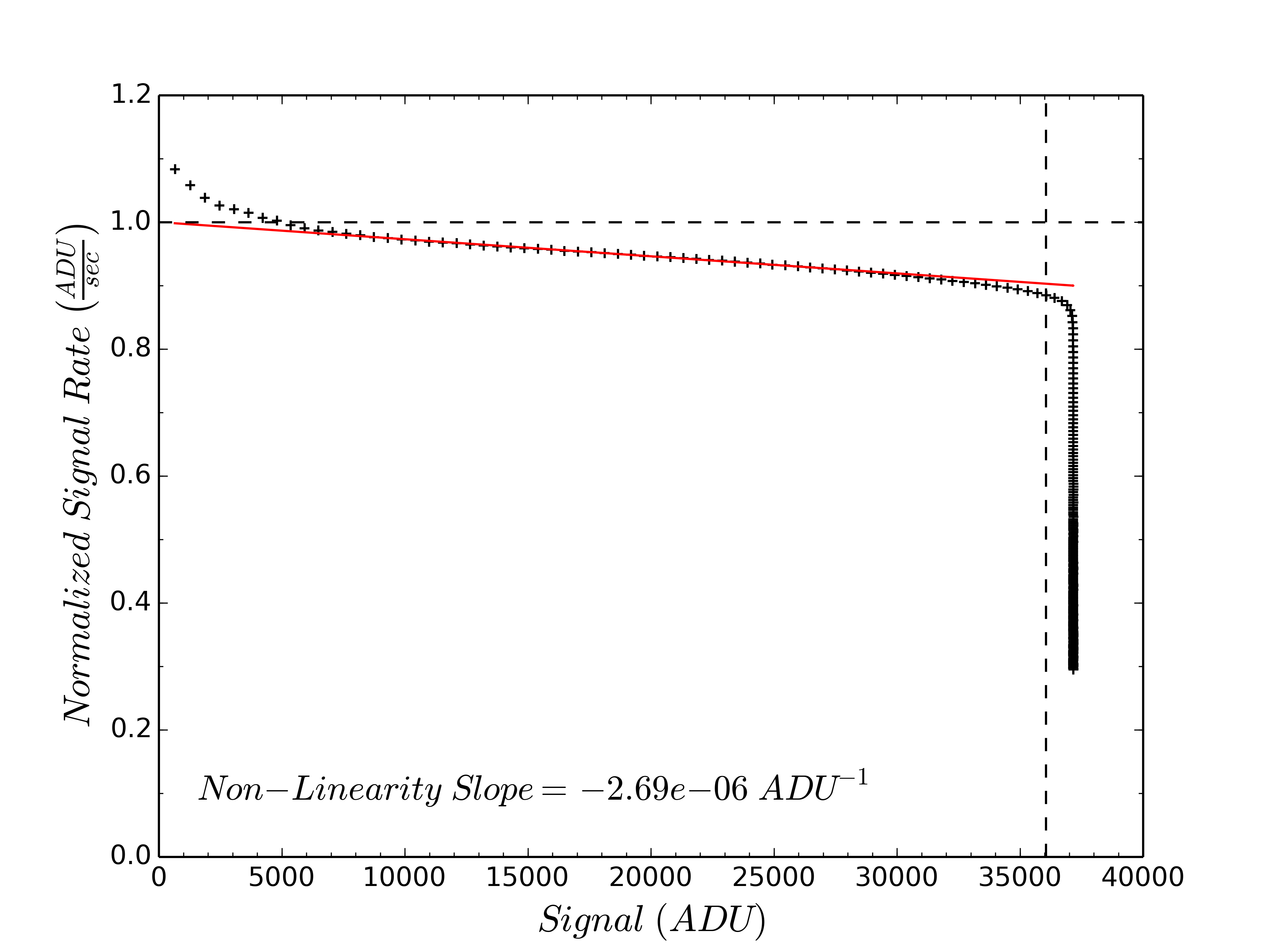}
\end{tabular}
\end{center}
\caption 
{ \label{fig:369_non-linearity_250mV}
Non-Linearity curve obtained from a 50X50 box average of pixels in H1RG-18369 with an applied bias of 250 mV at a temperature of 30 K. The signal rate was normalized and shifted such that the fitted line has a y-intercept of unity. The difference between a value of unity (horizontal dashed line) and the non-linearity curve is the fraction of the signal that was not collected due to the debiasing of the device. The vertical dashed line corresponds to the mean saturating signal at an actual bias of zero for the entire device. The rapid debiasing of the detector at low signals is due to tunneling dark currents.} 
\end{figure}

Figure \ref{fig:369_non-linearity_250mV} shows the non-linearity curve for H1RG-18369 with an applied bias of 250 mV, where at low signals the rate at which the diode debiases is much faster than is expected from a constant flux illumination. This quick debiasing is due to elevated dark current levels at high bias attributed to quantum mechanical tunneling. The effects of tunneling dark currents which are dominant at larger bias are further discussed in sections \ref{sect:IvsWell} and \ref{sect:dark current results}.

\subsection{Mux Glow}
\label{sect:mux glow}
In multiple occasions during the characterization of H1RG-18367 and H1RG-18369, a uniform elevated current was measured when taking data in the dark. This elevated current affected dark current at 150 mV (34 and 35 K), 250mV (33-35 K), and read noise at 30 K and 150 mV of applied bias for H1RG-18369. All of the dark current measurements for H1RG-18367 were affected by this elevated current.

A light leak in the test dewar has been ruled out because the anomalous dark current was not present in all data sets taken under similar conditions (without changes to the system setup). Figure \ref{fig:18369_Imodel} shows an instance where an elevated current affected the dark current and well depth data but was not present in the warm-up data (discussed in Sect. \ref{sect:I-T}) taken at a similar temperature and same applied bias. Dark current has also been ruled out since the dark current mechanisms studied here do not increase by a factor of 10 or more by increasing the temperature by 1 K for H1RG-18367 and -18369 (see Fig. \ref{fig:Ivstemp_150mV} or Tables \ref{tab:367 median I and well} and \ref{tab:369 median I and well}). Instead, we believe this current could be a glow from the unit cell FETs in the multiplexer. This effect will be referred to as ``mux glow'' hereafter when describing the data that were affected. Further work is required to evaluate this anomalous behavior.

\subsection{CDS Read Noise}
\label{sect:read noise}

The total noise per pixel was measured for two arrays, H1RG-18369 and H1RG-18509 at a temperature of 30 K and an applied bias of 150 mV. To compute the read noise, 64 SUTR sets (each SUTR consisting of 36 samples) were taken in the dark with an integration time of 5.5 seconds between each sample. From this, 64 Fowler-1 (CDS) images were created, where the total rms noise was calculated. The total CDS noise when measured in the dark is expected to be dominated by the read noise.

Shot noise from an elevated current measured for H1RG-18369 in this data set, most likely due to a ``mux glow", is subtracted in quadrature from the total CDS noise to obtain the CDS read noise. The current per pixel is calculated by fitting a line to each of the 64 SUTR curves taken in the dark, where the mean of the 64 fitted slopes is the current.

The mean dark current for H1RG-18509 was measured to be 0.8 $e^-/s$, whereas the elevated current for H1RG-18369 was 80 $e^-/s$, compared to the 0.7 $e^-/s$ measured in the data set taken to determine the dark current and well depth at the same temperature and bias (discussed in section \ref{sect:IvsWell}).

\begin{figure}
\begin{center}
\begin{tabular}{c}
\includegraphics[scale=0.6]{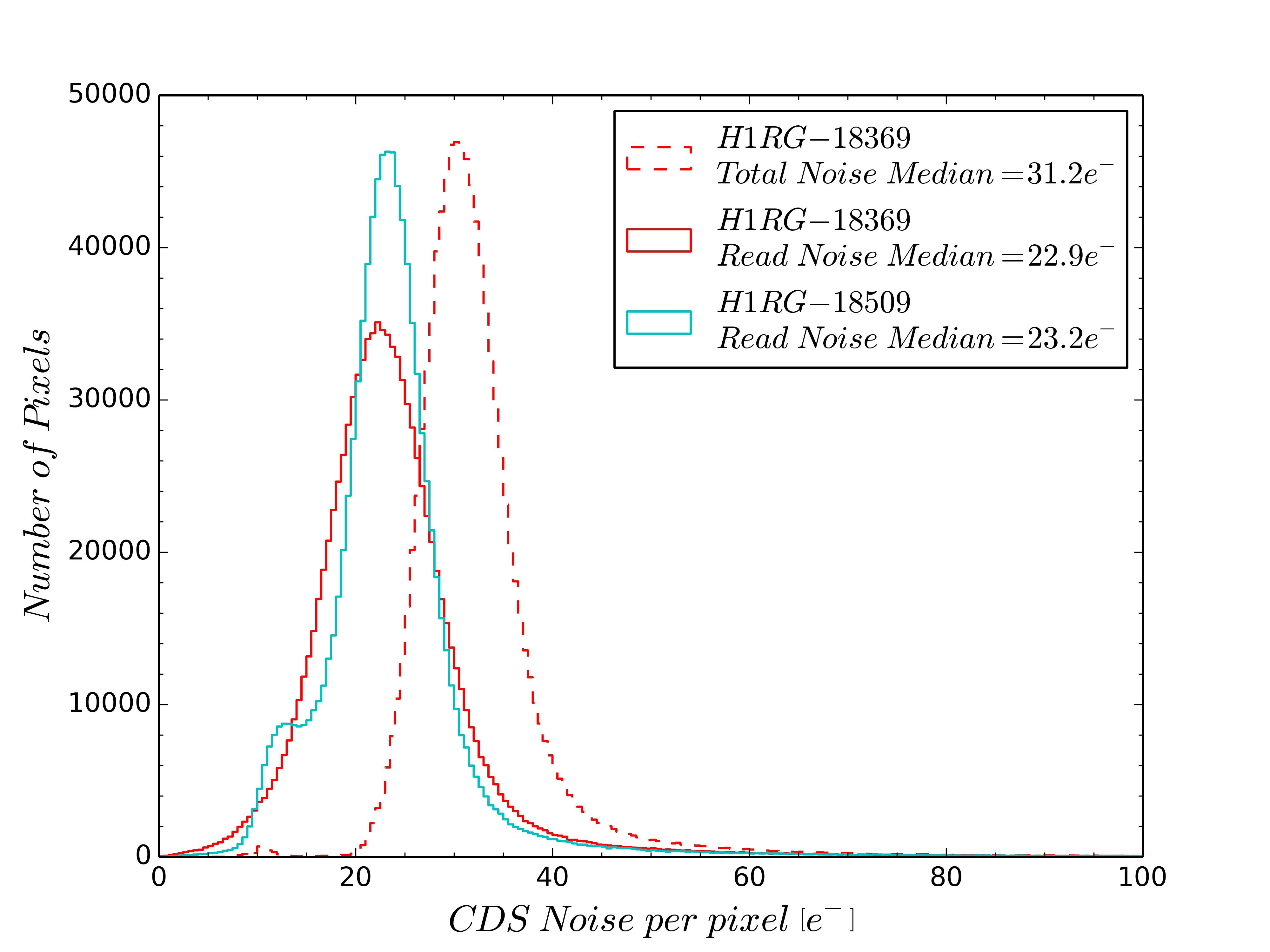}
\end{tabular}
\end{center}
\caption 
{ \label{fig:rms read noise}
Histogram of CDS read noise per pixel for H1RG-18369 and H1RG-18509 at a temperature of 30 K and applied bias of 150 mV. The CDS read noise was obtained by subtracting the contribution from dark current or a signal flux from the total noise in a CDS image. The total noise is plotted for H1RG-18369 to show the contribution from the ``mux glow'' to the noise of the detector. The dark current contribution from H1RG-18509 to the total noise was minimal, and did not exhibit a ``mux glow'', therefore we only present the read noise distribution for this array.} 
\end{figure}

Figure \ref{fig:rms read noise} shows the distribution of rms read noise per pixel for both arrays, where the median noise is 22.9 $e^-$ and 23.2 $e^-$ for H1RG-18369 and H1RG-18509 respectively. The majority of pixels with low noise approximately between 10 and 15 $e^-$ in the read noise distribution for H1RG-18509 and in the total noise distribution for H1RG-18369 are located in the ``picture frame'' region\cite{Rauscher13} of the arrays.

\subsection{Dark Current and Well Depth (Operability)}
\label{sect:IvsWell}

To measure the dark current and well depth, 200 samples-up-the-ramp in the dark were taken with an integration time of 5.8 seconds between each sample. Without resetting, we allow a radiative flux to uniformly saturate the array, where this saturating level includes a 10-20 mV forward bias contribution from the signal flux\cite{Bacon06}. To determine the actual zero bias saturation signal for each pixel, the array is then read out 200 additional times in the dark, allowing the array to debias to actual zero bias. The initial dark current is the measured slope at the beginning of the signal \textit{vs.} time (SUTR) curve. The difference between the pedestal frame and the average of the last twenty frames (taken when the array is at zero bias) then gives us the well depth of each pixel.

Well depth and dark current are presented here concurrently because the initial dark current of a pixel may be heavily dependent on the actual bias (well depth) across the pixel when it is first sampled (referred to as pedestal). Quantum tunneling dark currents are exponentially dependent on bias, so large tunneling dark currents can debias a pixel considerably between reset and pedestal frame, thus showing a low initial dark current but with a depleted well.

\subsubsection{H1RG-18367}
\label{sect:367_oper_discharge}

\begin{figure}
\begin{center}
\begin{tabular}{c}
\includegraphics[scale=0.5]{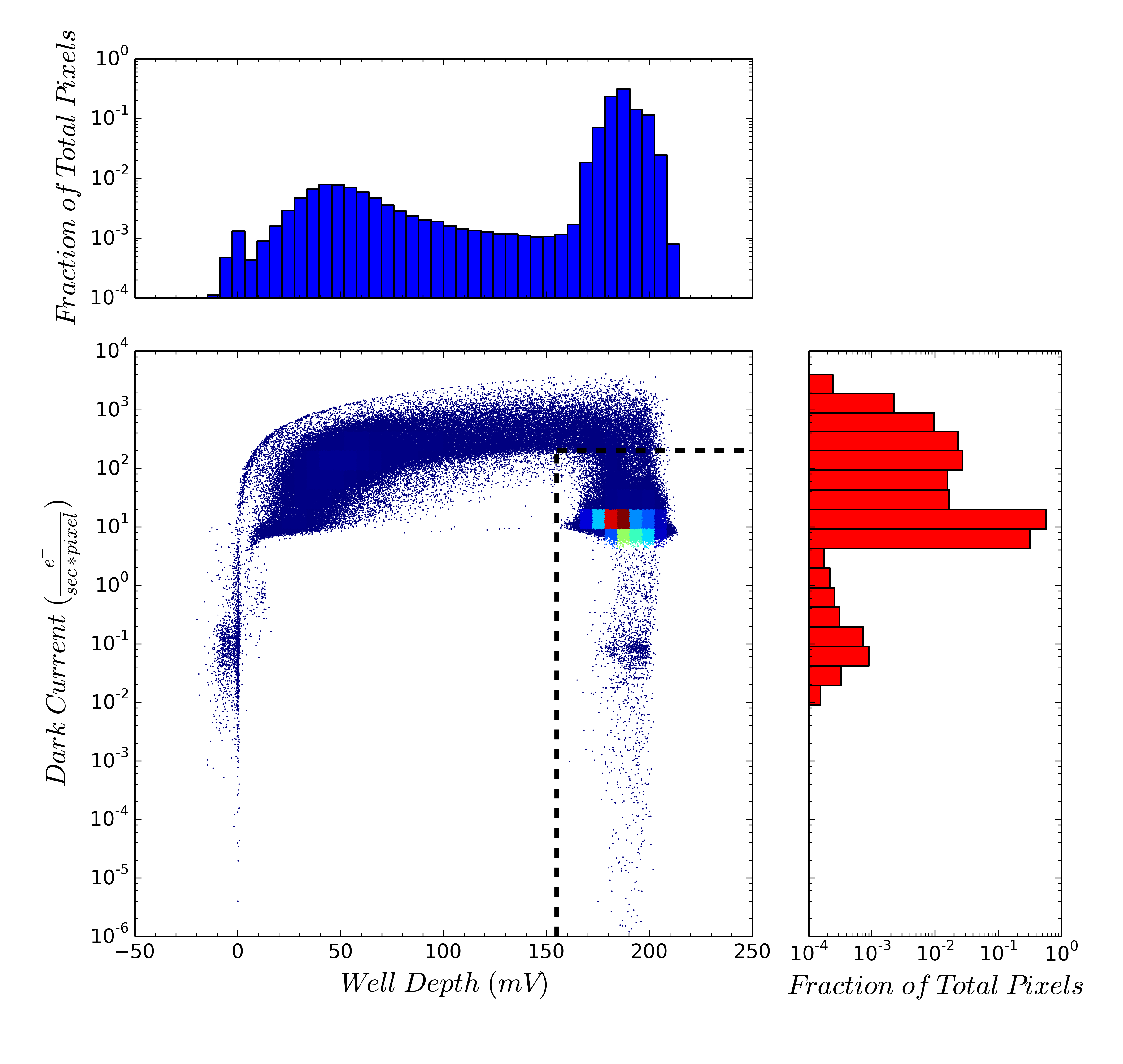}
\end{tabular}
\end{center}
\caption 
{ \label{fig:367_IvW}
Current in the dark \textit{vs.} well depth distribution per pixel for H1RG-18367 at a temperature of 28 K and an applied bias of 150 mV, showing operable pixels below and to the right of the dashed line. The well depth requirement of 155 mV ($\sim$ 41 $ke^-$) was chosen to include the majority of the good pixels, and we kept this requirement for higher temperatures with the same applied bias. Note that the well depth corresponds to the actual initial detector bias, ranging from 155 to 225 mV at the beginning of the integration among the operable pixels for this data set.}
\end{figure}

Figure \ref{fig:367_IvW} shows the distribution of initial dark current \textit{vs.} well depth per pixel for H1RG-18367 at a temperature of 28 K and applied bias of 150 mV, where the operable pixels are below and to the right of the dashed lines. The dark current distribution per pixel for this array peaks around 10 $e^-/s$ for an applied bias of 150 mV and a temperature of 28 K: we attribute this dark current to the ``mux glow'' discussed in section \ref{sect:mux glow} given its uniformity across the entire array, and lack of bias dependence. The ``mux glow'' was also observed at temperatures of 30 and 32 K for this array, with a median ``dark'' current of 130 and 156 $e^-/s$ respectively with a bias of 150 mV. This accounts for some of the loss of operability at 30 and 32 K \textit{c.f.} 28K. The percentage of operable pixels for different temperatures and applied biases are reported in Table \ref{tab:367_operability}, along with the well depth requirements. Median dark current and well depth measured for this array are shown in Table \ref{tab:367 median I and well}.

\begin{table}
\centering
\caption{Operability pixel percentage for H1RG-18367 with different applied biases and temperatures. Operability requirements include currents in the dark$^*$ below 200 $e^-/s$ and well depths greater than those indicated in the table.}
\begin{tabular}{cc|c|c|c|}
\cline{3-5}
                                   &                                        & T=28 K    & T=30 K    & T=32 K   \\ \hline
\multicolumn{1}{|c|}{\begin{tabular}[c]{@{}c@{}}Applied\\ Bias\end{tabular}} & \begin{tabular}[c]{@{}c@{}}Well Depth\\ Minimum\end{tabular}         & \multicolumn{3}{c|}{Operability (\%)} \\ \hline
\multicolumn{1}{|c|}{150 mV}       & \begin{tabular}[c]{@{}c@{}}155 mV\\ ($\sim$ 41 $ke^-$)\end{tabular} & 91.4    & 90.9    & 88.9   \\ \hline
\multicolumn{1}{|c|}{250 mV}       & \begin{tabular}[c]{@{}c@{}}255 mV\\ ($\sim$ 59 $ke^-$)\end{tabular} & 90.1    & 89.5    & 88.2   \\ \hline
\multicolumn{1}{|c|}{350 mV}       & \begin{tabular}[c]{@{}c@{}}310 mV\\ ($\sim$ 67 $ke^-$)\end{tabular} & 1.4     & 0.5     & 3.9    \\ \hline
\end{tabular}
\begin{threeparttable}
\label{tab:367_operability}
\begin{tablenotes}
\item[*] A glow from the mux affected all of the dark current measurements for this array. This ``mux glow'' was highly variable, with median current of approximately 10, 130, and 156 $e^-/s$ at temperatures of 28, 30, and 32 K respectively with an applied bias of 150 mV. 
\end{tablenotes}
\end{threeparttable}
\end{table}

\begin{table}
\centering
\caption{Median dark current and well depth for H1RG-18367. ``Mux glow" was present in all dark current measurements. With 350 mV of bias, the decrease of dark current with increasing temperature (due to increasing the band gap energy) is indicative of tunneling dark currents.}
\label{tab:367 median I and well}
\begin{tabular}{c|c|c|c|}
\cline{2-4}
 & T=28 K & T=30 K & T=32 K \\ \hline
\rowcolor[HTML]{C0C0C0} 
\multicolumn{1}{|c|}{\cellcolor[HTML]{C0C0C0}\begin{tabular}[c]{@{}c@{}}Applied\\ Bias\end{tabular}} & \multicolumn{3}{c|}{\cellcolor[HTML]{C0C0C0}\begin{tabular}[c]{@{}c@{}}Median Dark Current $\left(e^-/sec\right)$\\ Median Well Depth $\left(ke^-, mV\right)$\end{tabular}} \\ \hline
\multicolumn{1}{|c|}{} & 10 & 130 & 156 \\ \cline{2-4} 
\multicolumn{1}{|c|}{\multirow{-2}{*}{150 mV}} & 50, 186 & 50, 185 & 48, 180 \\ \hline
\rowcolor[HTML]{C0C0C0} 
\multicolumn{1}{|c|}{\cellcolor[HTML]{C0C0C0}} & 12 & 116 & 139 \\ \cline{2-4} 
\rowcolor[HTML]{C0C0C0} 
\multicolumn{1}{|c|}{\multirow{-2}{*}{\cellcolor[HTML]{C0C0C0}250 mV}} & 67, 287 & 67, 286 & 66, 281 \\ \hline
\multicolumn{1}{|c|}{} & 379 & 359 & 321 \\ \cline{2-4} 
\multicolumn{1}{|c|}{\multirow{-2}{*}{350 mV}} & 82, 379 & 82, 379 & 82, 376 \\ \hline
\end{tabular}
\end{table}

\begin{figure}
\begin{center}
\begin{tabular}{c}
\includegraphics[scale=0.5]{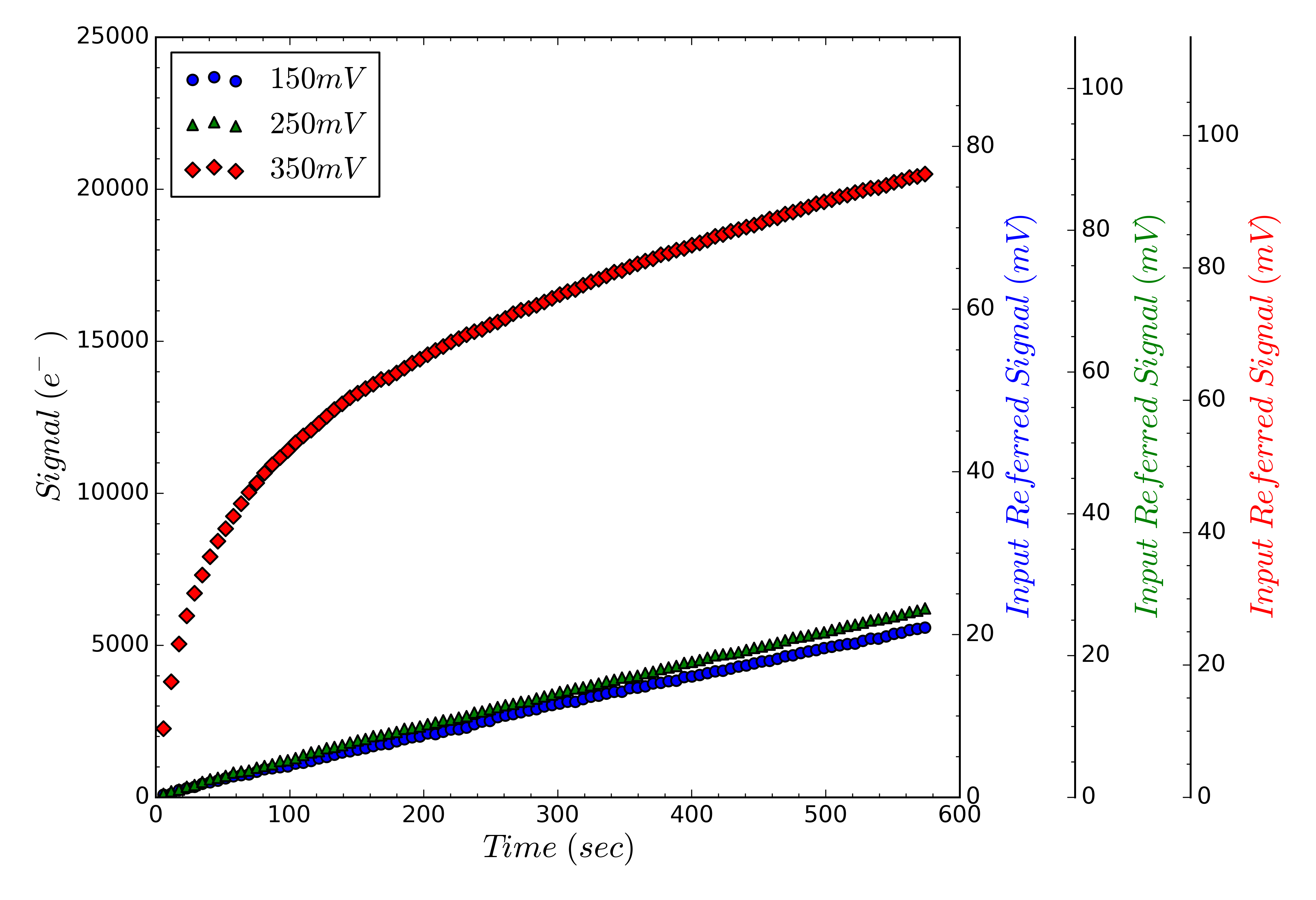}
\end{tabular}
\end{center}
\caption 
{ \label{fig:367_sutr}
Discharge history for pixel [532, 468] from H1RG-18367 at a temperature of 28K and at applied reverse bias of 150, 250, and 350mV in the dark. On the left y-axis scale we show the signal in electrons, while the right y-scale axis shows the input referred signal in mV. The three different scales on the right correspond to the different diode capacitance measured (from left to right) at 150, 250, and 350 mV of applied biases. This pixel is inoperable at 350mV of applied bias, since the initial dark current exceeds the 200 $e^-/s$ cutoff.} 
\end{figure} 

Figure \ref{fig:367_sutr} shows the discharge history of a pixel in the dark at three different applied biases for this device. The initial larger curvature observed in the signal with an initial applied bias of 350 mV is a consequence of quantum tunneling dark currents which are exponentially dependent on bias. As the pixel debiases, tunneling dark currents decrease and the dark current approaches the constant behavior with bias expected from steady ``mux glow''.

The major limiting factor in the operability of these arrays at larger bias and low temperatures has been tunneling dark currents, and it is much more evident at biases of 350 mV since trap-to-band and especially band-to-band tunneling are strongly dependent on the applied bias \cite{Wu97,Bailey98,Bacon10}. At higher applied bias, most of the pixels have dark currents exceeding 200 $e^-/s$, giving very low operabilities. The trap-to-band dark current is highly variable from pixel to pixel, but the band-to-band tunneling affects all of the pixels in the array.

\begin{figure}
\begin{center}
\begin{tabular}{c}
\includegraphics[scale=0.2]{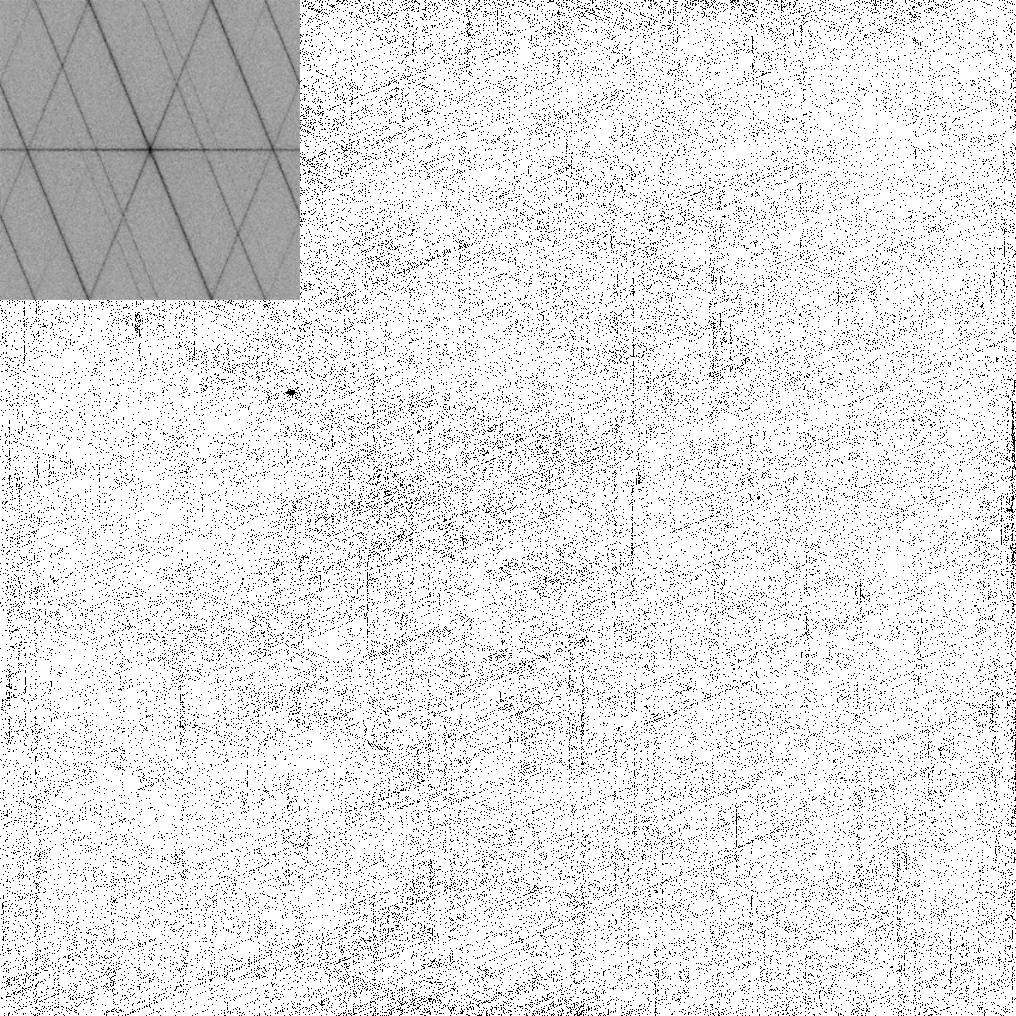}
\end{tabular}
\end{center}
\caption 
{ \label{fig:367_oper}
Operability map for H1RG-18367 at a temperature of 28 K and applied bias of 150 mV, where inoperable pixels are shown in black. The (log magnitude) FFT of the operability map is shown in the upper-left corner.} 
\end{figure} 

\begin{figure}
\begin{center}
\begin{tabular}{c}
\includegraphics[scale=0.3]{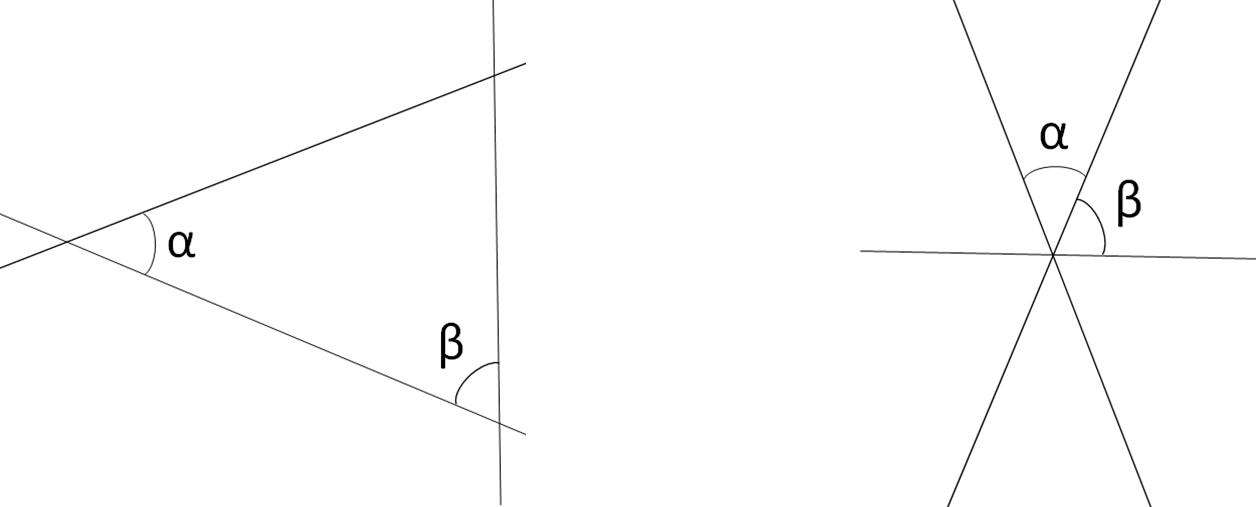}
\\
\hspace{0.5cm} (a) \hspace{5.5cm} (b)
\end{tabular}
\end{center}
\caption 
{ \label{fig:cross-hatching lines and angles}
Diagram of the three cross-hatching lines that were observed in the operability map (a), and in the (log magnitude) FFT of the operability map (b). The angles, $\alpha = \ang{44.4}$ and $\beta = \ang{67.9}$, were determined by applying the Hough transform on the FFT image.} 
\end{figure} 

The inoperability at low temperatures and low applied bias ($\sim$ 28 K and 150 mV respectively) appears to be dominated by trap-to-band tunneling currents, and not by band-to-band tunneling. This effect can be seen in the operability map for all four arrays (Fig.\ref{fig:367_oper} for H1RG-18367) where a set of high dark current and/or low well depth, hence inoperable pixels form a cross-hatching pattern caused by the slight lattice mismatch between HgCdTe and the CdZnTe substrate. The lattice mismatch may cause the formation of misfit dislocations along the intersection of the $\{111\}$ slip planes in zinc blende crystals and the growth plane of these arrays\cite{Martinka2001,Chang2008,Ayers07}. The fast Fourier transform (FFT) of the operability map on the upper left corner of Fig. \ref{fig:367_oper} shows the cross-hatching lines distinctly, but rotated by 90 degrees.

The pattern consists of a set of three lines which have been identified by other authors\cite{Martinka2001,Chang2008} to lie parallel to the $\left[\overline{2}31\right]$, $\left[\overline{2}13\right]$, and $\left[01\overline{1}\right]$ directions. The angles between the three cross-hatching lines (shown in Fig. \ref{fig:cross-hatching lines and angles}) were calculated by applying the Hough transform to the FFT image. The angles found in this work, $\alpha = \ang{44.4}$ and $\beta = \ang{67.9}$, agree with the angles found in Martinka et al.\cite{Martinka2001} and Chang et al.\cite{Chang2008}. Shapiro et al.\cite{Shapiro18} see the same pattern along the three cross-hatching directions as QE variation on sub-pixel scales on a 2.5 $\mu m$ cutoff HgCdTe array, in addition to a cluster of high dark current pixels that lie along only one of the cross-hatching directions.

\subsubsection{H1RG-18508}
\label{sect:508_oper_discharge}

\begin{table}
\centering
\caption{Operability pixel percentage for H1RG-18508 with different applied biases and temperatures. Operability requirements include dark currents below 200 $e^-/s$ and well depths greater than those indicated in the table.}
\label{tab:508 operability}
\begin{tabular}{cc|c|c|c|c|c|c|c|}
\cline{3-9}
                                   &                                                                     & T=28 K & T=30 K & T=32 K & T=33 K & T=34 K & T=35 K & T=36 K \\ \hline
\multicolumn{1}{|c|}{\begin{tabular}[c]{@{}c@{}}Applied\\ Bias\end{tabular}} & \begin{tabular}[c]{@{}c@{}}Well Depth\\ Minimum\end{tabular}         & \multicolumn{7}{c|}{Operability (\%)}                        \\ \hline
\multicolumn{1}{|c|}{150 mV}       & \begin{tabular}[c]{@{}c@{}}155 mV\\ ($\sim$ 41 $ke^-$)\end{tabular} & 93.7   & 93.5   & 93.2   & 92.9   & 92.0   & 61.5   & 63.7   \\ \hline
\multicolumn{1}{|c|}{250 mV}       & \begin{tabular}[c]{@{}c@{}}245 mV\\ ($\sim$ 58 $ke^-$)\end{tabular} & 89.4   & 91.5   & 91.4   & 91.0   & 89.9   & 81.1   & 67.2   \\ \hline
\end{tabular}
\end{table}

\begin{table}
\centering
\caption{Median dark current and well depth for H1RG-18508.}
\label{tab:508 median I and well}
\begin{tabular}{c|c|c|c|c|c|c|c|}
\cline{2-8}
 & T=28 K & T=30 K & T=32 K & T=33 K & T=34 K & T=35 K & T=36 K \\ \hline
\rowcolor[HTML]{C0C0C0} 
\multicolumn{1}{|c|}{\cellcolor[HTML]{C0C0C0}\begin{tabular}[c]{@{}c@{}}Applied\\ Bias\end{tabular}} & \multicolumn{7}{c|}{\cellcolor[HTML]{C0C0C0}\begin{tabular}[c]{@{}c@{}}Median Dark Current $\left(e^-/sec\right)$\\ Median Well Depth $\left(ke^-, mV\right)$\end{tabular}} \\ \hline
\multicolumn{1}{|c|}{} & 0.5 & 1.4 & 6 & 14 & 32 & 65 & 143 \\ \cline{2-8} 
\multicolumn{1}{|c|}{\multirow{-2}{*}{150 mV}} & 46, 177 & 45, 173 & 45, 171 & 44, 170 & 44, 169 & 41, 157 & 42, 160 \\ \hline
\rowcolor[HTML]{C0C0C0} 
\multicolumn{1}{|c|}{\cellcolor[HTML]{C0C0C0}} & 57 & 39 & 34 & 39 & 51 & 71 & 144 \\ \cline{2-8} 
\rowcolor[HTML]{C0C0C0} 
\multicolumn{1}{|c|}{\multirow{-2}{*}{\cellcolor[HTML]{C0C0C0}250 mV}} & 65, 275 & 64, 271 & 63, 269 & 63, 268 & 63, 267 & 60, 254 & 60, 253 \\ \hline
\multicolumn{1}{|c|}{} & 780 & 764 & 730 & 713 & 693 & 510 & 279 \\ \cline{2-8} 
\multicolumn{1}{|c|}{\multirow{-2}{*}{350 mV}} & 73, 328 & 74, 331 & 76, 337 & 76, 341 & 77, 344 & 76, 341 & 74, 331 \\ \hline
\end{tabular}%
\end{table}

\begin{figure}
\begin{center}
\begin{tabular}{c}
\includegraphics[scale=0.5]{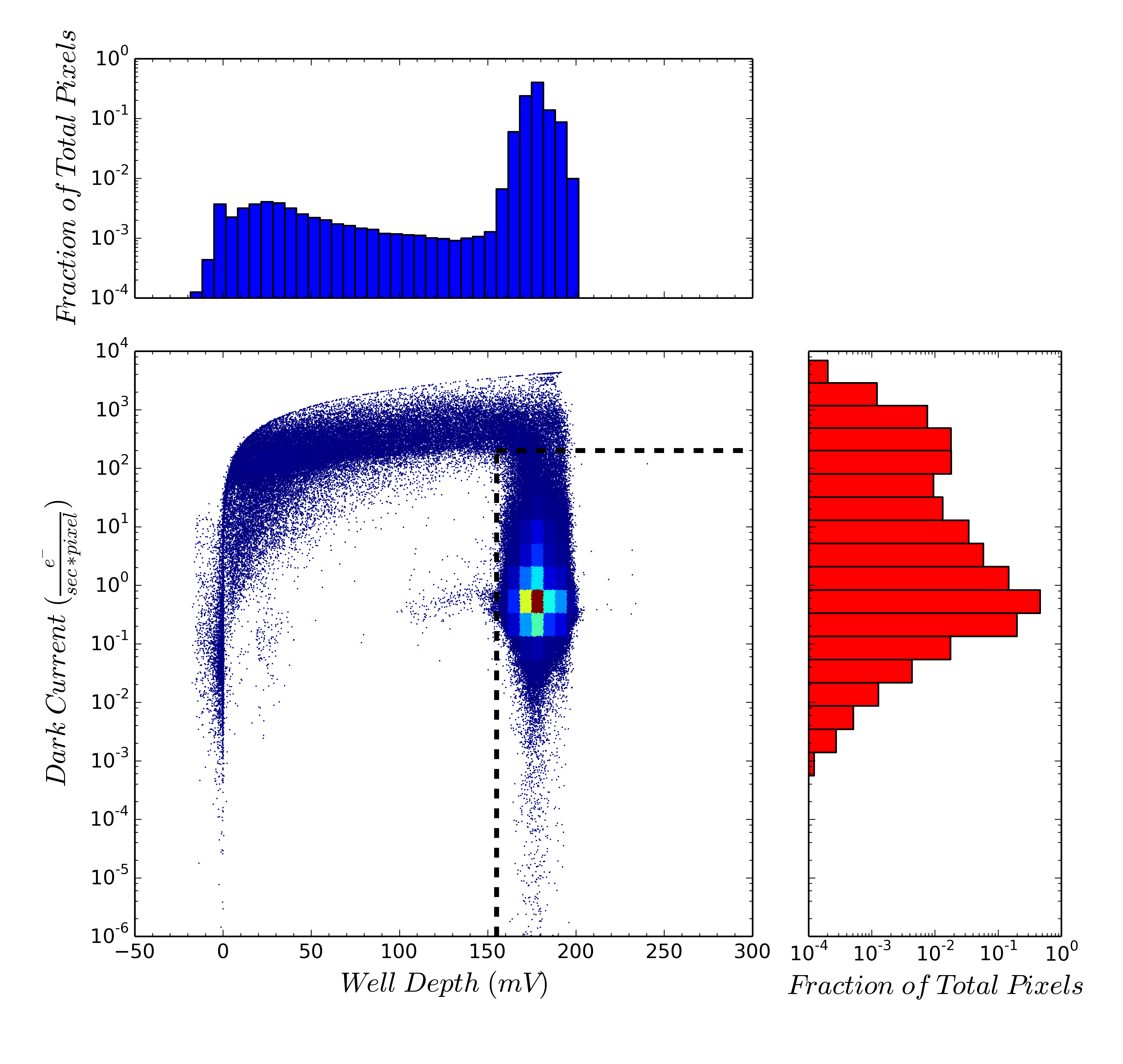}
\end{tabular}
\end{center}
\caption 
{ \label{fig:508_IvW}
Dark Current \textit{vs.} well depth distribution per pixel for H1RG-18508 with an applied bias of 150 mV at a temperature of 28 K. Operable pixels lie below and to the right of the dashed line. The detector bias at the beginning of integration is similar to that of array H1RG-18367 (\textit{c.f.} Fig.\ref{fig:367_IvW}).} 
\end{figure}

H1RG-18508 has a higher operability (Table \ref{tab:508 operability}) at higher temperatures than its counterpart H1RG-18367 (both were grown and processed in the same way, and both have similar cutoff wavelengths), and ``mux glow'' was not detected for this array. The dark current \textit{vs.} well depth distribution for this array at 28 K and an applied bias of 150 mV is shown in Fig. \ref{fig:508_IvW}, where the majority of pixels had dark currents below 1 $e^-/s$. Median dark current and well depths are presented in Table \ref{tab:508 median I and well}.

\begin{figure}
\begin{center}
\begin{tabular}{c}
\includegraphics[scale=0.5]{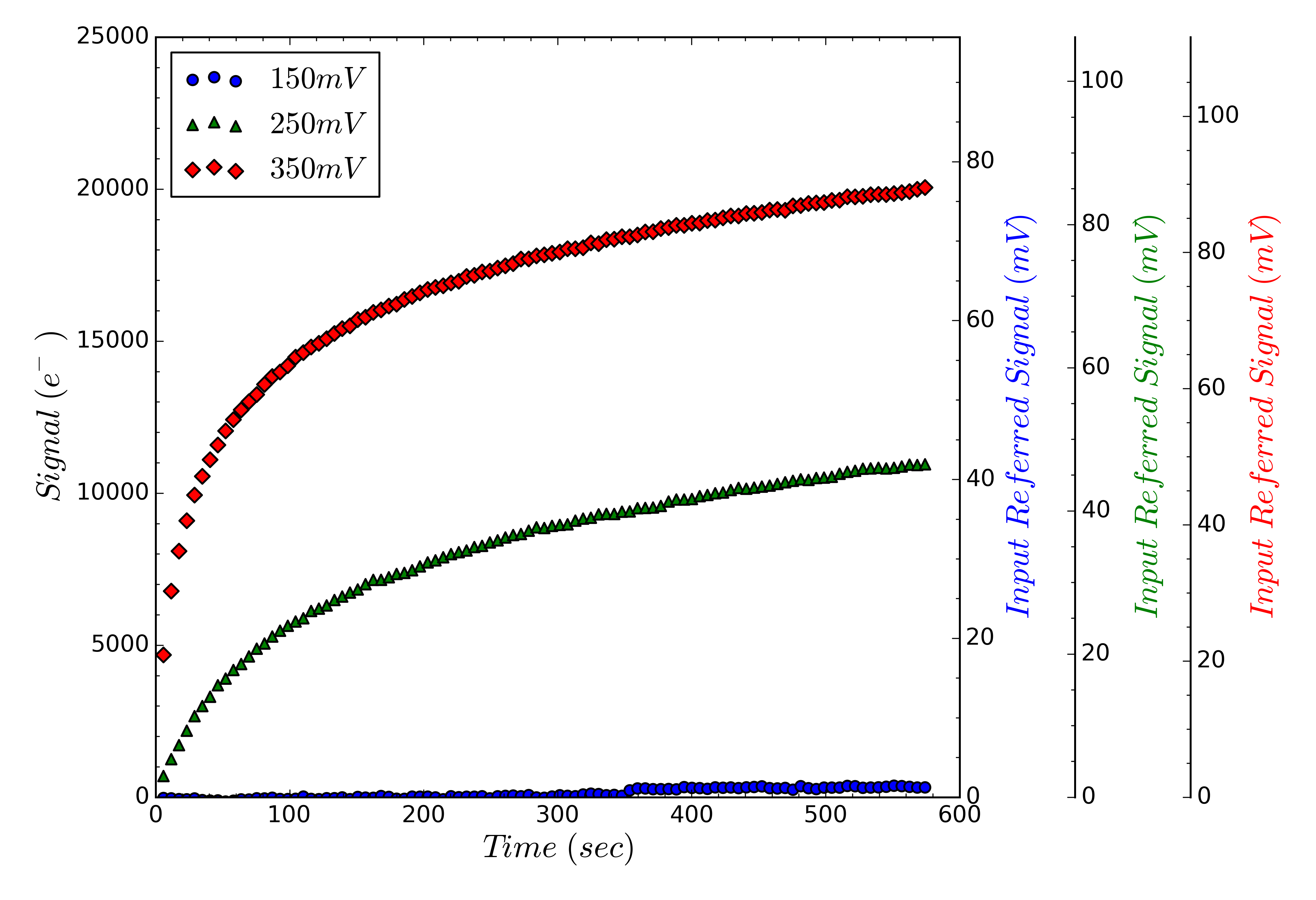}
\end{tabular}
\end{center}
\caption 
{ \label{fig:508_sutr}
Discharge history for pixel [543, 424] from H1RG-18508 at a temperature of 28K in the dark. The initial dark currents (and well depths) for three SUTR curves at 150, 250, and 350 mV are: 0.4 $e^-/s$ (181 mV), 120 $e^-/s$ (281 mV), and 807 $e^-/s$ (326 mV) respectively.} 
\end{figure}

The operabilities at 350 mV of applied bias are below 1\% at all temperatures, and have therefore been omitted from the operability table. The inoperability at larger applied bias is evidence of tunneling dark currents. The large curvature in the dark signal \textit{vs.} time curve is observed with increasing bias in Fig. \ref{fig:508_sutr}, which is a hallmark of tunneling dark currents. The SUTR data obtained with an applied bias of 150 mV are linear as is expected if thermal dark currents (diffusion and G-R, which are a weak function of bias), a light leak, or a mux glow are dominating the dark current as the pixel debiases.

\subsubsection{H1RG-18369}

\label{sect:369_oper_discharge}
\begin{table}
\centering
\caption{Operability pixel percentage for H1RG-18369 with different applied biases and temperatures. Operability requirements include dark currents below 200 $e^-/s$ and well depths greater than those indicated in the table.}
\label{tab:369 operability}
\begin{tabular}{cc|c|c|c|c|c|c|}
\cline{3-8}
                                   &                                        & T=28 K & T=30 K & T=32 K & T=33K  & T=34 K & T=35 K \\ \hline
\multicolumn{1}{|c|}{\begin{tabular}[c]{@{}c@{}}Applied\\ Bias\end{tabular}} & \begin{tabular}[c]{@{}c@{}}Well Depth\\ Minimum\end{tabular} & \multicolumn{6}{c|}{Operability (\%)}                    \\ \hline
\multicolumn{1}{|c|}{150 mV}       & \begin{tabular}[c]{@{}c@{}}155 mV\\ ($\sim$ 37 $ke^-$)\end{tabular} & 93.0 & 92.8 & 92.7 & 92.6 & 0.0$^*$  & 0.0$^*$  \\ \hline
\multicolumn{1}{|c|}{250 mV}       & \begin{tabular}[c]{@{}c@{}}255 mV\\ ($\sim$ 57 $ke^-$)\end{tabular} & 91.5 & 91.3 & 91.1 & 0.0$^*$  & 0.0$^*$  & 0.0$^*$  \\ \hline
\end{tabular}
\begin{threeparttable}
\begin{tablenotes}
\item[*] The ``mux glow'' affected the noted operabilities for this array, increasing the median dark current by a factor of $\sim$100 when the temperature was increased from 33 to 34 K with an applied bias of 150 mV, and by a factor of $\sim$ 43 when increasing the temperature from 32 to 33 K with an applied bias of 250 mV, effectively making the entire array inoperable.
\end{tablenotes}
\end{threeparttable}
\end{table}

\begin{table}
\centering
\caption{Median dark current and well depth for H1RG-18369. Dark currents marked with an asterisk were affected by a ``mux glow".}
\label{tab:369 median I and well}
\begin{tabular}{c|c|c|c|c|c|c|}
\cline{2-7}
 & T=28 K & T=30 K & T=32 K & T=33 K & T=34 K & T=35 K \\ \hline
\rowcolor[HTML]{C0C0C0} 
\multicolumn{1}{|c|}{\cellcolor[HTML]{C0C0C0}\begin{tabular}[c]{@{}c@{}}Applied\\ Bias\end{tabular}} & \multicolumn{6}{c|}{\cellcolor[HTML]{C0C0C0}\begin{tabular}[c]{@{}c@{}}Median Dark Current $\left(e^-/sec\right)$\\ Median Well Depth $\left(ke^-, mV\right)$\end{tabular}} \\ \hline
\multicolumn{1}{|c|}{} & 0.5 & 0.7 & 2.4 & 5 & 481* & 698* \\ \cline{2-7} 
\multicolumn{1}{|c|}{\multirow{-2}{*}{150 mV}} & 43, 182 & 42, 177 & 41, 174 & 43, 181 & 41, 173 & 40, 169 \\ \hline
\rowcolor[HTML]{C0C0C0} 
\multicolumn{1}{|c|}{\cellcolor[HTML]{C0C0C0}} & 24 & 17 & 14 & 604* & 600* & 791* \\ \cline{2-7} 
\rowcolor[HTML]{C0C0C0} 
\multicolumn{1}{|c|}{\multirow{-2}{*}{\cellcolor[HTML]{C0C0C0}250 mV}} & 64, 283 & 63, 278 & 62, 275 & 62, 275 & 62, 273 & 61, 269 \\ \hline
\multicolumn{1}{|c|}{} & 730 & 682 & 616 & 989* & 930* & 957* \\ \cline{2-7} 
\multicolumn{1}{|c|}{\multirow{-2}{*}{350 mV}} & 78, 355 & 78, 357 & 79, 359 & 80, 364 & 80, 364 & 79, 361 \\ \hline
\end{tabular}%
\end{table}

\begin{figure}
\begin{center}
\begin{tabular}{c}
\includegraphics[width=\textwidth]{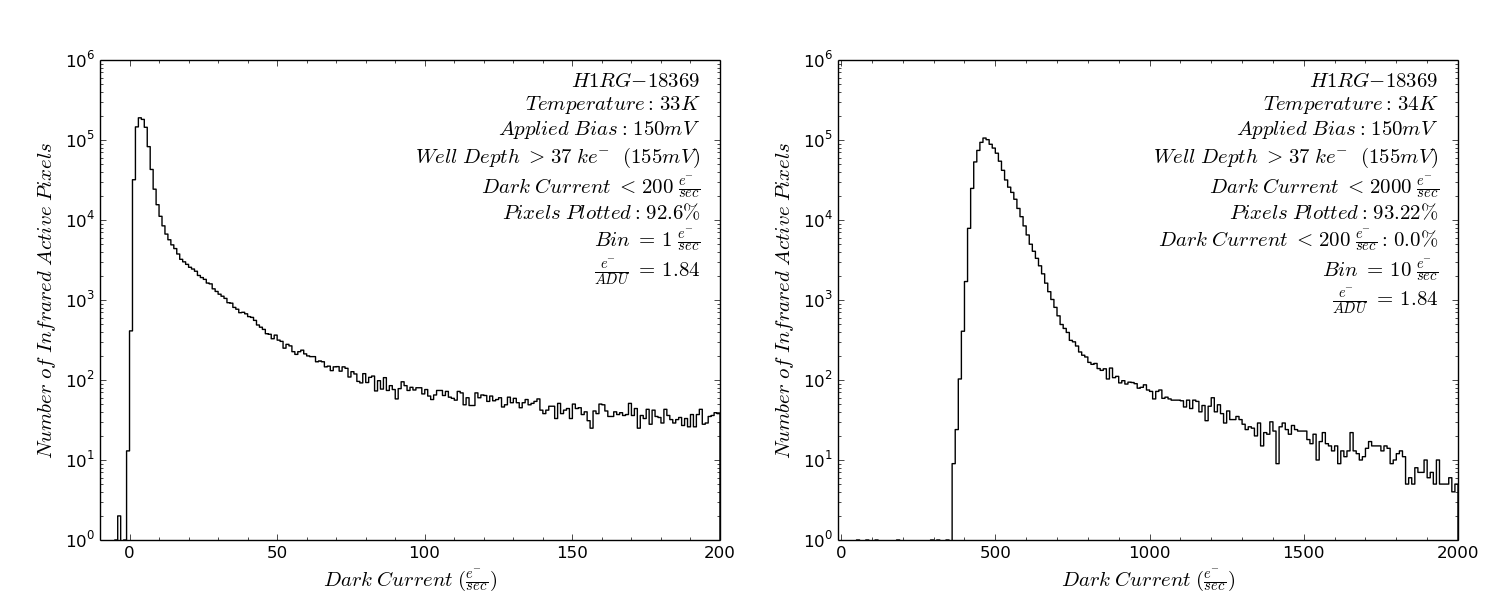}
\\
(a) \hspace{7.4cm} (b)
\end{tabular}
\end{center}
\caption 
{ \label{fig:369_I_hist_33K_34K}
Dark current distribution for pixels with well depths greater than 37 $ke^-$ at temperatures of 33 K (a) and 34 K (b), both with an applied bias of 150 mV. The increase in current by a factor of $\sim$100 by increasing the temperature by one degree Kelvin is most probably due to ``mux glow''.} 
\end{figure}

This array has a similar dark current and well depth distribution as H1RG-18508, where the majority of the pixels achieved dark current levels less than 1 $e^-/s$ at a temperature of 28 K and an applied bias of 150 mV. Table \ref{tab:369 operability} shows the operability for this array, where the unusual drop in operability when increasing the temperature by one degree with an applied bias of 150 mV and 250 mV is most likely due to ``mux glow''. Table \ref{tab:369 median I and well} shows the median dark current and well depth for this array, where the increase in median dark current by a factor of $\sim$100 at a temperature of 34 K and 150 mV of bias cannot be explained by any of the known dark current mechanisms. Additionally, the decrease in dark current with increasing temperature, for an applied bias of 350 mV from 28-32 K, is consistent with tunneling dark currents. If the increase in dark current at 33 K was due to thermal currents overtaking tunneling currents, we would expect the dark current to continue to rise with increasing temperature, and this is not the case. Figure \ref{fig:369_I_hist_33K_34K} shows the dark current distribution for this array at 33 and 34 K with an applied bias of 150 mV.

Furthermore, despite the experimental design used to reduce tunneling dark currents, this array also exhibited low operabilities at 350 mV ($<$ 1\%) as did the standard growth arrays (H1RG-18367 and 18508).

\subsubsection{H1RG-18509}
\label{sect:509_oper_discharge}

\begin{table}
\centering
\caption{Operable pixel percentage for H1RG-18509 with different applied biases and temperatures. Operability requirements include dark currents below 200 $e^-/s$ and well depths greater than those indicated in the table.}
\label{tab:509 operability}
\begin{tabular}{cc|c|c|c|c|c|c|}
\cline{3-8}
                                   &                                                                                   & T=28 K & T=30 K & T=32 K & T=34 K & T=35 K & T=36 K \\ \hline
\multicolumn{1}{|c|}{\begin{tabular}[c]{@{}c@{}}Applied\\ Bias\end{tabular}} & \begin{tabular}[c]{@{}c@{}}Well Depth\\ Minimum\end{tabular}                       & \multicolumn{6}{c|}{Operability (\%)}                    \\ \hline
\multicolumn{1}{|c|}{150 mV}       & \begin{tabular}[c]{@{}c@{}}155 mV\\ ($\sim$ 38 $ke^-$)\end{tabular} & 95.9 & 95.9 & 95.8 & 95.7 & 94.7 & 85.5 \\ \hline
\multicolumn{1}{|c|}{250 mV}       & \begin{tabular}[c]{@{}c@{}}255 mV\\ ($\sim$ 59 $ke^-$)\end{tabular} & 94.7 & 94.6 & 94.5 & 94.0 & 92.7 & 44.7 \\ \hline
\multicolumn{1}{|c|}{350 mV}       & \begin{tabular}[c]{@{}c@{}}355 mV\\ ($\sim$ 75 $ke^-$)\end{tabular} & 90.3 & 92.2 & 93.1 & 92.3 & 90.8 & 19.4 \\ \hline
\end{tabular}
\end{table}

\begin{table}
\centering
\caption{Median dark current and well depth for H1RG-18509.}
\label{tab:509 median I and well}
\begin{tabular}{c|c|c|c|c|c|c|}
\cline{2-7}
 & T=28 K & T=30 K & T=32 K & T=34 K & T=35 K & T=36 K \\ \hline
\rowcolor[HTML]{C0C0C0} 
\multicolumn{1}{|c|}{\cellcolor[HTML]{C0C0C0}\begin{tabular}[c]{@{}c@{}}Applied\\ Bias\end{tabular}} & \multicolumn{6}{c|}{\cellcolor[HTML]{C0C0C0}\begin{tabular}[c]{@{}c@{}}Median Dark Current $\left(e^-/sec\right)$\\ Median Well Depth $\left(ke^-, mV\right)$\end{tabular}} \\ \hline
\multicolumn{1}{|c|}{} & 0.2 & 0.7 & 3.5 & 19 & 41 & 82 \\ \cline{2-7} 
\multicolumn{1}{|c|}{\multirow{-2}{*}{150 mV}} & 44, 182 & 43, 177 & 43, 174 & 42, 174 & 42, 173 & 40, 163 \\ \hline
\rowcolor[HTML]{C0C0C0} 
\multicolumn{1}{|c|}{\cellcolor[HTML]{C0C0C0}} & 0.3 & 0.8 & 3.8 & 20 & 43 & 76 \\ \cline{2-7} 
\rowcolor[HTML]{C0C0C0} 
\multicolumn{1}{|c|}{\multirow{-2}{*}{\cellcolor[HTML]{C0C0C0}250 mV}} & 65, 283 & 64, 278 & 63, 276 & 63, 275 & 63, 274 & 59, 255 \\ \hline
\multicolumn{1}{|c|}{} & 1.8 & 1.8 & 4 & 19 & 42 & 67 \\ \cline{2-7} 
\multicolumn{1}{|c|}{\multirow{-2}{*}{350 mV}} & 81, 385 & 80, 380 & 80, 378 & 80, 377 & 80, 376 & 74, 348 \\ \hline
\end{tabular}%
\end{table}

To mitigate the effects of the tunneling dark currents, Teledyne developed several modifications to its baseline NEOCam design, designated as designs 1 and 2, shown in Table \ref{tab:QE and cutoff}. H1RG-18509 was from the Design-2 lot split and outperformed the other LW13 arrays in terms of operability at all applied biases, but especially at 350 mV, and at higher temperatures (Table \ref{tab:509 operability}). Table \ref{tab:509 median I and well} shows the median dark current and well depths measured for this array at different temperatures and applied bias. This array has a slightly shorter wavelength cutoff compared with the standard growth arrays. 

\begin{figure}
\begin{center}
\begin{tabular}{c}
\includegraphics[scale=0.5]{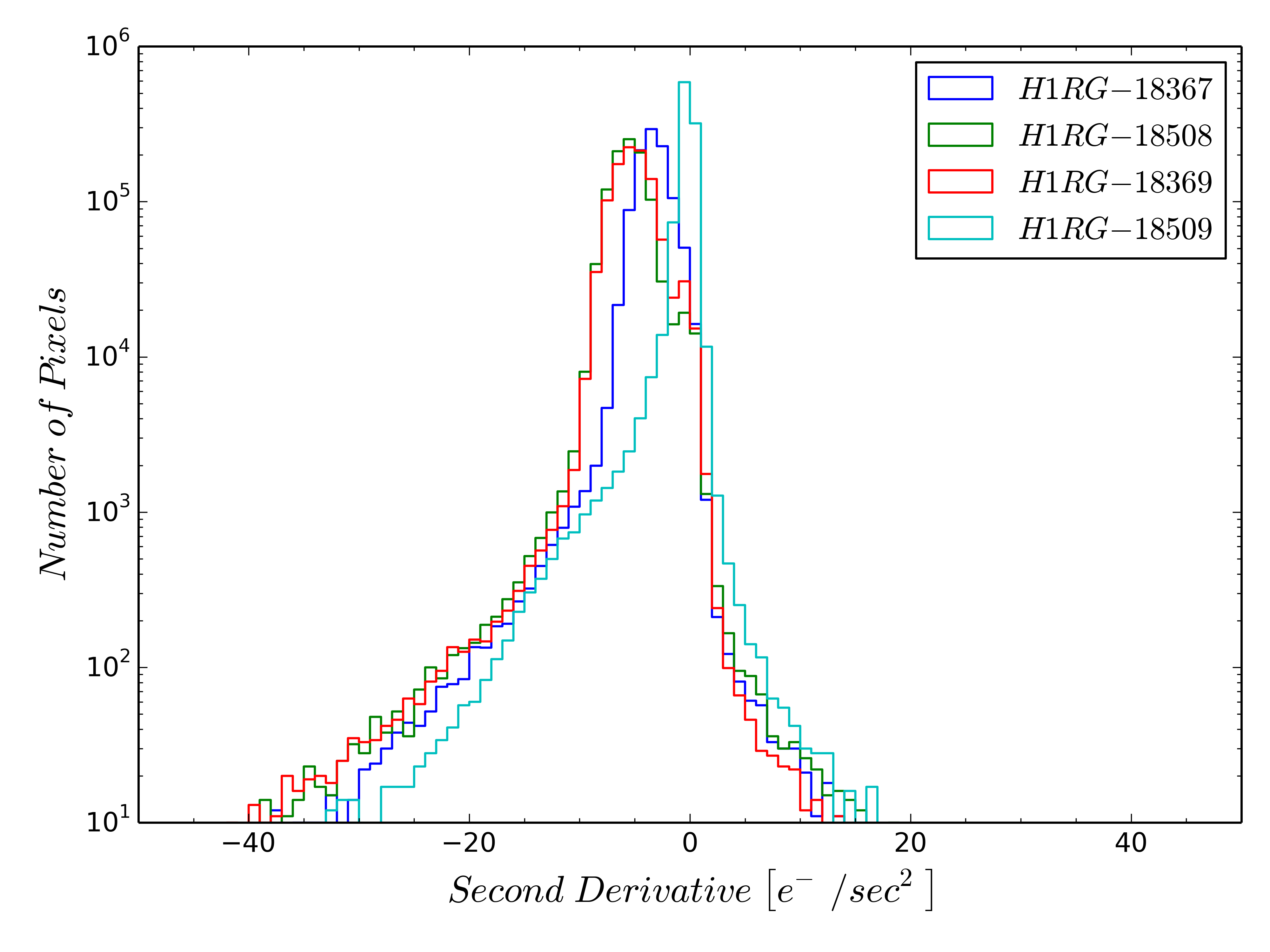}
\end{tabular}
\end{center}
\caption 
{ \label{fig:curvature_350mV_28K}
Histogram of the initial curvature from SUTR curves for all LW13 arrays at 28 K and an applied bias of 350mV.} 
\end{figure}

Design 2 had a positive effect on reducing the quantum tunneling currents as is readily apparent in curves of the time dependence of charge collected in the dark, where most of the individual SUTR curves for H1RG-18509 did not exhibit highly curved discharge behavior such as observed for the other arrays. Figure \ref{fig:curvature_350mV_28K} shows the histogram of curvature values (initial second derivative of the SUTR curves) per pixel, for all LW13 arrays with an applied bias of 350 mV. The peak of the histogram near zero for H1RG-18509 demonstrates a nearly linear behavior in the charge collected over time for the majority of pixels, in contrast to the behavior of the other three arrays shown by the peaks of their respective histograms shifted towards larger negative values. At these higher biases, band-to-band tunneling current becomes dominant and this affects all pixels.

\section{Dark Current Model Fits}
\label{sect:dark current results}

The effects of tunneling dark currents were shown in sections \ref{sect:367_oper_discharge} - \ref{sect:509_oper_discharge}, and comparing our measurements to theory will allow us to assess the degree to which each of the tunneling components is dominant and how we can mitigate these effects further as we continue to extend the cutoff wavelength for the second phase of this project.

Given the different dependences on temperature and bias for the dark current mechanisms, we fit the thermal currents, which are strongly temperature dependent, to dark current \textit{vs.} temperature (I-T) data, while the tunneling currents are fit to dark current \textit{vs.} bias (I-V) data given their strong dependence on bias.

\subsection{Dark Current \textit{vs.} Temperature}
\label{sect:I-T}

I-T data consist of two different data sets: (1) the initial dark current per pixel which was obtained at the stable temperatures reported in the operability tables in Sections \ref{sect:367_oper_discharge} - \ref{sect:509_oper_discharge}, and (2) the warm-up data taken when the liquid helium in our dewar runs out. Four full array frames are read after resetting the array in SUTR mode, immediately followed by an array reset and reading four sub-array (32 rows, all columns) frames in SUTR mode. This data-taking process continued until the temperature reached 77 K. The dark current was then obtained by subtracting the pedestal from the following three frames, taking the average signal from the pedestal subtracted frames and dividing by the average integration time.

The integration time between the full array data frames is 5.8 seconds, and 0.2 seconds for the sub-array frames. At lower temperatures, the amount of charge collected by pixels in 0.2 seconds is on the order of the read-noise ($\sim$ 23 $e^-$); we therefore use the 5.8 second integration time data to form the I-T curve at these lower temperatures. As the temperature increases, so does the dark current, saturating the full array frames at a temperature of $\sim$ 40 K. At a point before the full frames saturate we used the sub-array frames which saturate at a higher temperature ($\sim$ 50 K) because of the shorter integration time.

\subsection{Dark Current \textit{vs.} Bias}
\label{sect:I-V}

SUTR data sets taken to measure the dark current and well depth are used to obtain I-V curves. The dark current is obtained by taking the difference between sequential data points along the dark signal vs. time curve, and dividing by the time interval between those points, while the actual bias across the diode is calculated by subtracting the input-referred signal from the well depth level (see the upper curves in Figs. \ref{fig:367_sutr} and \ref{fig:508_sutr}). Only partial I-V curves can be obtained from the dark SUTR curves since we do not completely debias the devices in the dark. To reduce the effect of thermal dark currents on the I-V curve, tunneling dark currents are fit to the lowest temperature I-V curve (28 K).

The I-V curve for the 150 mV SUTR set is not used since the data for reasonably good pixels typically span less than 5 mV of bias. Instead, only the initial dark current and well depth are used as the lowest bias data point. The same is done for H1RG-18509 for the three applied biases since most pixels discharge less than 5 mV in the dark signal \textit{vs.} time data.

\subsection{Fitting Process and Caveats}
\label{sect: Fitting process and caveats}

\begin{figure}
\begin{center}
\begin{tabular}{c}
\includegraphics[width=\textwidth]{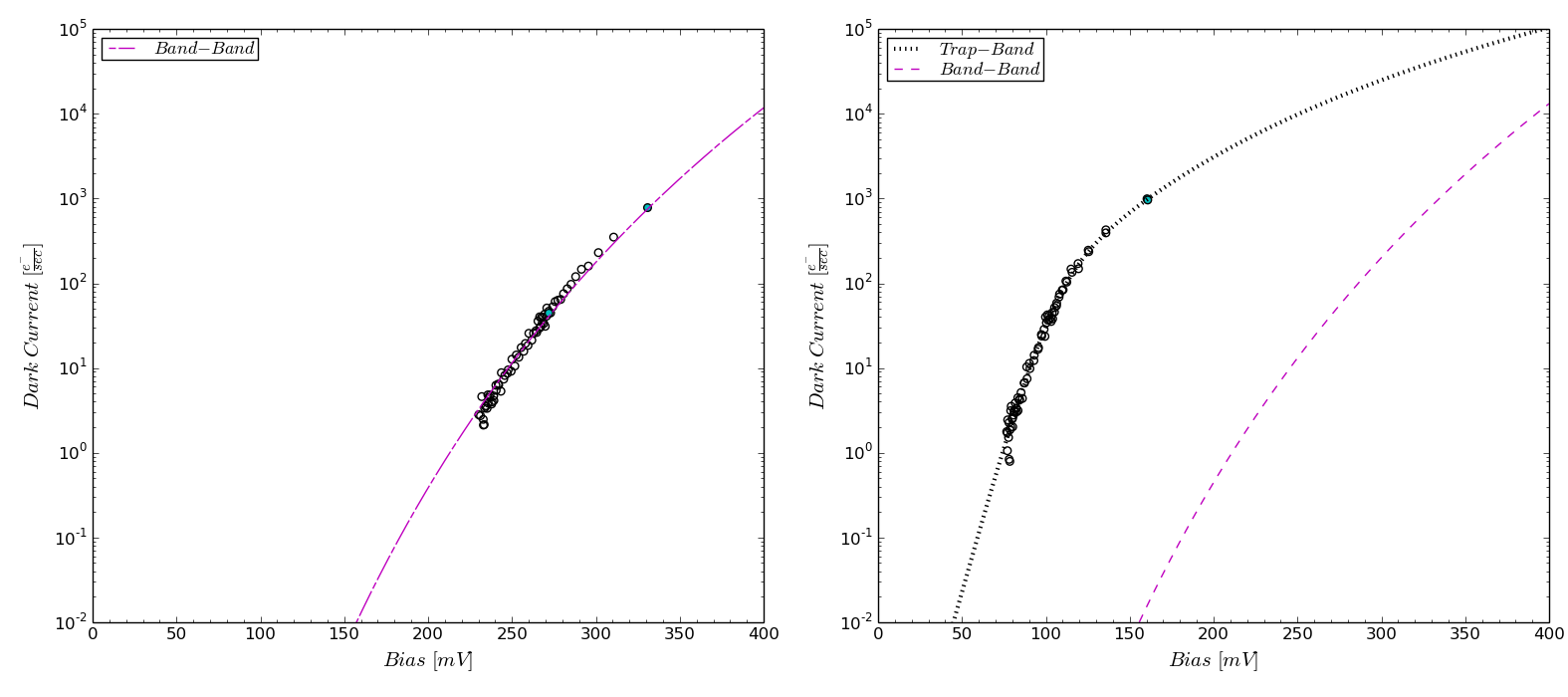}
\\
(a) \hspace{7.4cm} (b)
\end{tabular}
\end{center}
\caption 
{ \label{fig:508_betafit}
Dark current \textit{vs.} bias at a temperature of 28 K for (a) an operable and (b) inoperable (at 150 mV of applied bias) pixel in H1RG-18508.} 
\end{figure}

Equations \ref{eq:band-to-band} and \ref{eq:trap-to-band} in Section \ref{sect:tunneling} show that tunneling currents are exponentially dependent on band gap and electric field in the junction region. The I-V curves of several operable pixels at large applied bias ($>$ 200 mV) matched the trend of a fit to band-to-band tunneling, where the only parameter to fit is $E_g^{3/2}/E \equiv \beta$. If indeed these pixels (or the majority) are dominated by band-to-band tunneling, this would allow us to get the most accurate estimate of the parameter $\beta$ since other dark current mechanisms have several other parameters to fit, which can lead to a non-unique set of fitted parameters. Figure \ref{fig:508_betafit} (a) shows an example of an operable pixel in H1RG-18508. Inoperable pixels required the addition of trap-to-band tunneling to fit the I-V data. Figure \ref{fig:508_betafit} (b) shows an inoperable pixel in H1RG-18508, where the $\beta$ parameter of the operable nearest neighbor pixels used to fit band-to-band tunneling current is shown to be orders of magnitude below the measured dark current and hardly contributes to the fit.

Unlike band-to-band tunneling where $\beta$ is the only unknown parameter, trap-to-band tunneling requires the fitting of five additional parameters (see Sect. \ref{sect:tunneling}). Furthermore, the fitted trap-to-band tunneling heavily depends on the parameters' initial guess, which may not be unique.

Therefore, only band-to-band is initially fitted to the larger applied bias I-V data of operable pixels to estimate the value of $\beta$. The following step to model the dark current was to then fit the thermal dark currents (diffusion and G-R) to the higher temperature I-T data. Lastly, the test dewar light leak or ``mux glow'' is fitted, along with trap-to-band tunneling if necessary.

When estimating the value of $\beta$ in H1RG-18367 and H1RG-18509, an additional constant current was added to the I-V data along with band-to-band tunneling (similar to that shown on Fig. \ref{fig:18367_Imodel} on individual pixels) because the large bias data was not completely dominated by band-to-band tunneling. In the case of H1RG-18367, the ``glow'' from the multiplexer (discussed in sections \ref{sect:mux glow} and \ref{sect:367_oper_discharge}) has a considerable effect on the I-V data up to 300 mV of bias, while H1RG-18509 is dominated by a possible $<$ 1 $e^-/s$ light leak in the test dewar and G-R up to a bias of $\sim$350 mV. Since G-R current does not have a large bias dependence in the range of interest, the light leak plus the G-R current in H1RG-18509 were modeled as a constant current when estimating $\beta$.

\begin{figure}
\begin{center}
\begin{tabular}{c}
\includegraphics[scale=0.7]{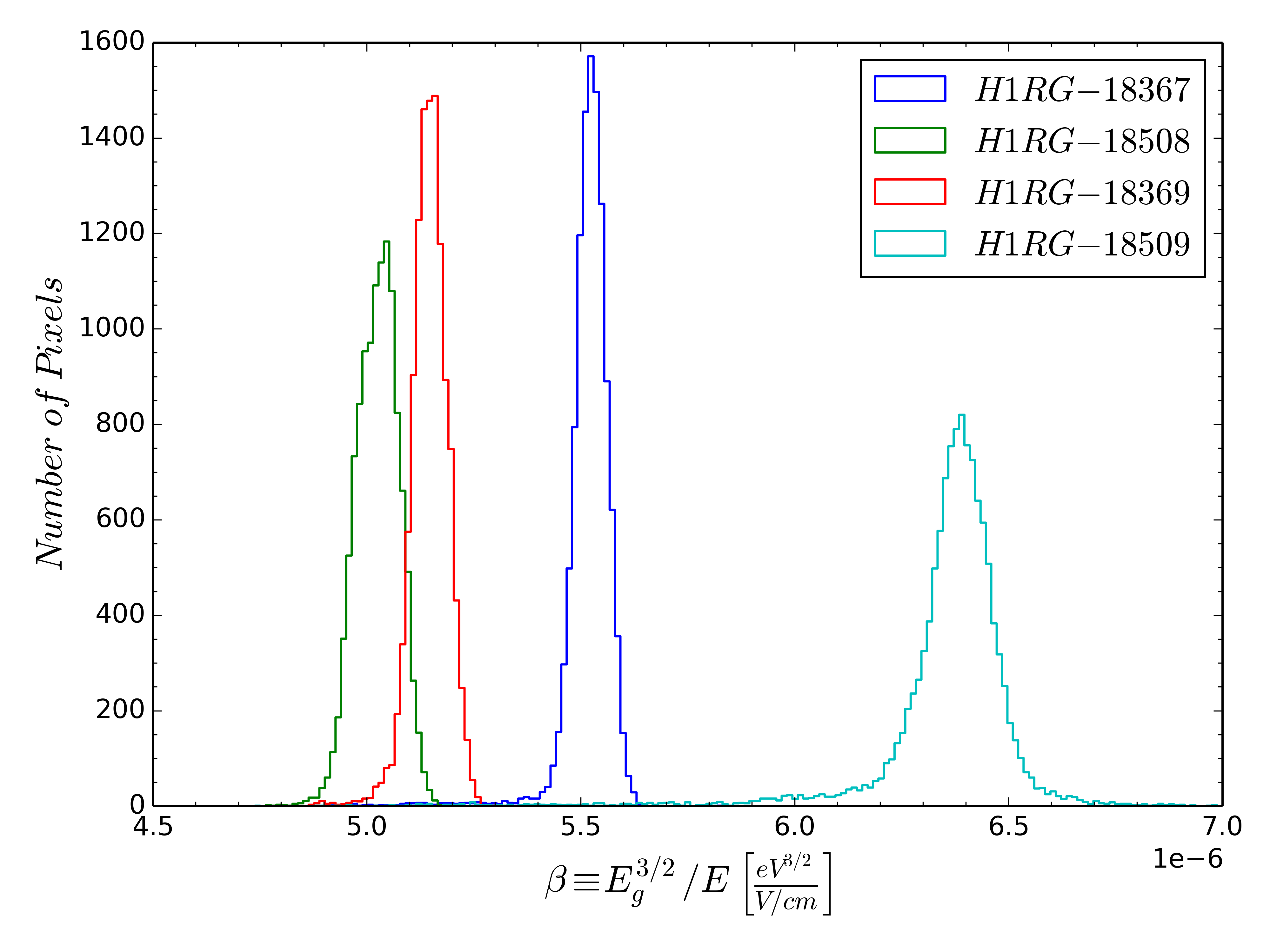}
\end{tabular}
\end{center}
\caption 
{ \label{fig:beta fit}
Fitted $\beta$ value distribution, with an actual reverse bias of 300 mV, to operable pixels in the central 32 rows (for which we have warm-up data) and 400 columns of the array.}
\end{figure}

Figure \ref{fig:beta fit} shows the distribution of fitted $\beta$ values to a 32$\times$400 pixel region in the four arrays presented here.

\subsection{Dark Current Model Results}
\label{sect:dark model results}

\begin{figure}
\begin{center}
\begin{tabular}{c}
\includegraphics[scale=0.4]{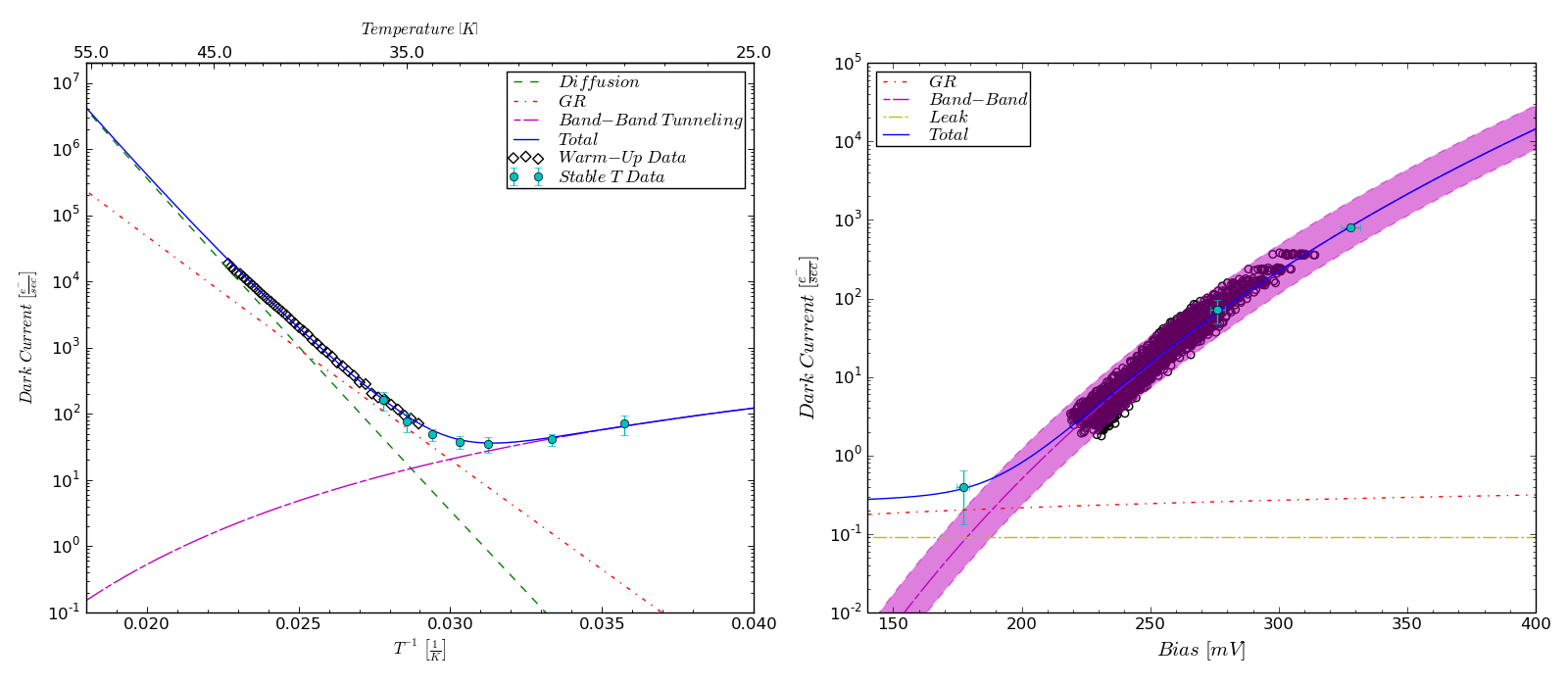}
\\
(a) \hspace{7.4cm} (b)
\end{tabular}
\end{center}
\caption 
{ \label{fig:18508_Imodel}
(a) Dark current \textit{vs.} temperature (mean actual back-bias of 276 mV) averaged data of 36 operable pixels in H1RG-18508. The error bars correspond to $\pm$ one standard deviation of the mean in the dark current values for the averaged pixels. (b) Individual dark current \textit{vs.} bias (at 28 K) data curves (empty circle data points) for the same 36 pixels shown in (a). The solid cyan circle data points are the average of the initial dark current and actual bias for the 36 pixels. The shaded region in (b) corresponds to the band-to-band tunneling calculated from the $\beta$ value $\pm$ two standard deviations from the mean, while the single band-to-band tunneling curve corresponds to that of the mean $\beta$ value of the 36 pixels.} 
\end{figure}

Figure \ref{fig:18508_Imodel} shows the the dark current Arrhenius plot at a detector bias of 276 mV and dark current \textit{vs.} bias (at a temperature of 28 K) for 36 operable pixels in H1RG-18508. The I-T data shown is the average of all 36 pixels, where the error bars on the filled circles correspond to the standard deviation of the mean in the dark current values for the averaged pixels. The individual I-V data curves for each of the 36 pixels are displayed along with the mean initial dark current and actual bias (filled circles) for the 36 pixels corresponding to 150, 250, and 350 mV of applied bias. The shaded region's upper and lower bounds coincide with the band-to-band tunneling calculated from the $\beta$ values that are two standard deviations away from the mean of the fitted $\beta$ value distribution in Fig. \ref{fig:beta fit}, while the single band-to-band tunneling curve is calculated from the mean $\beta$ value of the 36 pixels.

The I-V data for the 36 pixels in Fig. \ref{fig:18508_Imodel} (b) follow the trend of band-to-band tunneling with slightly different $\beta$ values at biases $>$ 200 mV. The dark current models fitted to the I-T data shows that, up to a temperature of $\sim$32 K, the dark current of operable pixels is dominated by band-to-band tunneling. At higher temperatures G-R and diffusion currents are the dominating components.

\begin{figure}
\begin{center}
\begin{tabular}{c}
\includegraphics[scale=0.4]{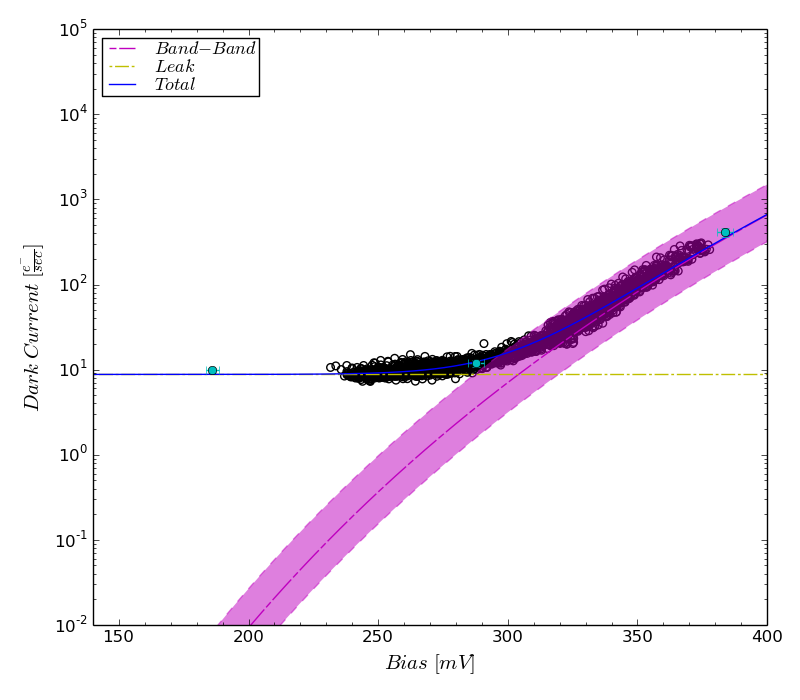}
\end{tabular}
\end{center}
\caption 
{ \label{fig:18367_Imodel}
Dark current \textit{vs.} bias data curves for 36 pixels which where operable at a temperature of 28 K and applied bias of 250 mV in H1RG-18367 at a temperature of 28 K. The constant leak current that was fitted corresponds to the ``mux glow''.} 
\end{figure}

\begin{figure}
\begin{center}
\begin{tabular}{c}
\includegraphics[scale=0.4]{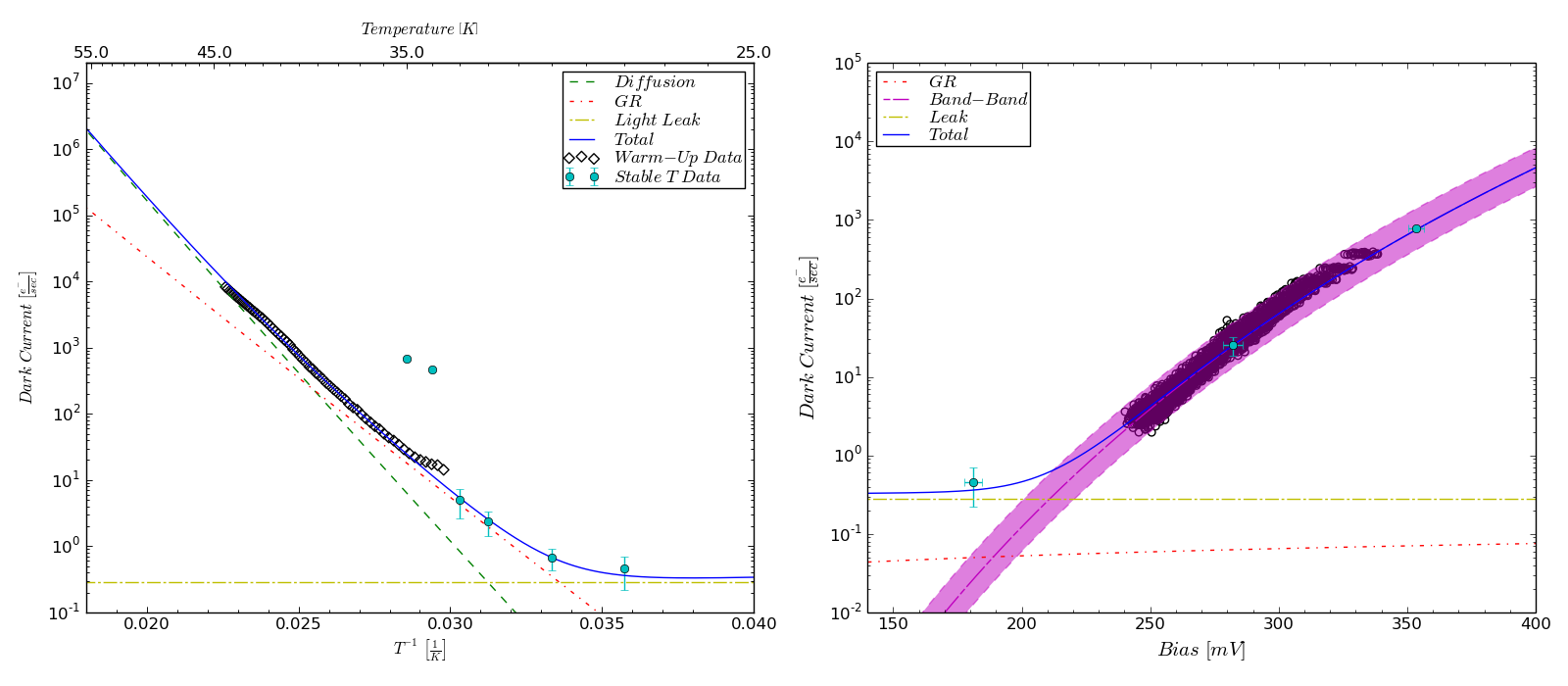}
\\
(a) \hspace{7.4cm} (b)
\end{tabular}
\end{center}
\caption 
{ \label{fig:18369_Imodel}
(a) Dark current \textit{vs.} temperature (applied bias of 150 mV) averaged data of 36 operable pixels in H1RG-18369. The mean actual bias of 181 mV, corresponding to the 28 K stable data point, was used to fit the dark current models. (b) Individual dark current \textit{vs.} bias (at 28 K) data curves for the same 36 pixels shown in (a). The 33 and 34 K stable temperature data points in (a) were affected by a ``mux glow'', increasing the dark current above the expected value from thermal dark currents.} 
\end{figure}

\begin{figure}
\begin{center}
\begin{tabular}{c}
\includegraphics[scale=0.4]{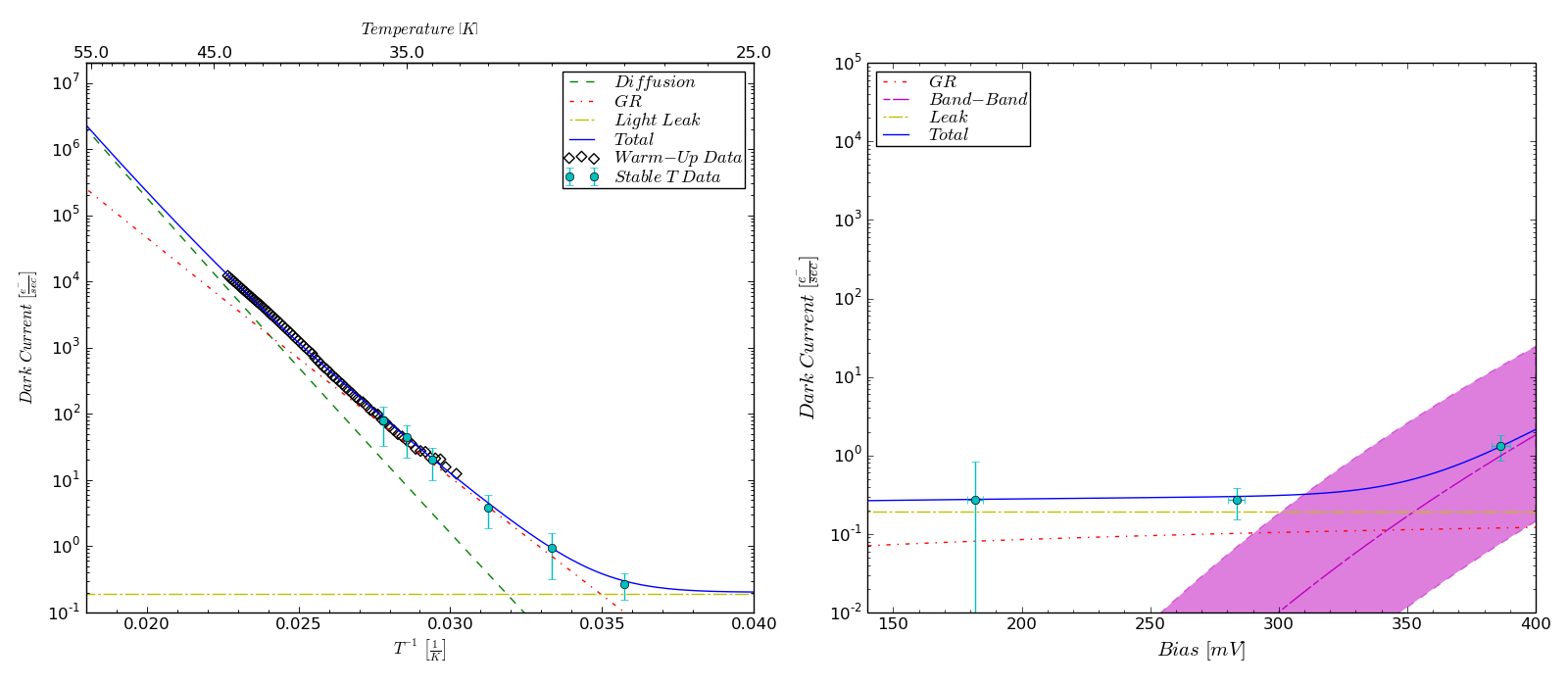}
\end{tabular}
\end{center}
\caption 
{ \label{fig:18509_Imodel}
(a) Dark current \textit{vs.} temperature (applied bias of 250 mV) and (b) dark current \textit{vs.} bias (at 28 K) averaged data of 50 operable pixels in H1RG-18509. In (a), the mean actual bias of 284 mV, corresponding to the 28 K stable data point, was used to fit the dark current models.} 
\end{figure}

Similar to H1RG-18508, the majority of the pixels in the other three arrays (Fig. \ref{fig:18367_Imodel}-\ref{fig:18509_Imodel}) that were operable at a temperature of 28 K and 250 mV of applied bias were dominated by band-to-band tunneling at larger biases

The observed ``mux glow'' in H1RG-18367 is large enough to dominate the dark current up to a bias of $\sim$ 300 mV (see Fig. \ref{fig:18367_Imodel}), and in the linear behavior of the SUTR curve for the pixel in Fig. \ref{fig:367_sutr}. Thermal dark currents were not fitted to this array as warm-up data are not available.

The warm-up data for H1RG-18369 were obtained with an applied bias of 150 mV, where the I-T data in Fig.\ref{fig:18369_Imodel} (a) show that at 28 K, the dark current is limited by a possible 0.3 $e^-/s$ light leak in the test dewar. The 34 and 35 K stable data points, which were affected by the ``mux glow'', do not follow the behavior expected from any of the dark current mechanisms, further confirming that the anomalous increase of dark current from 33 to 34 K is not due to thermal currents. The ``mux glow'' was not present in the warm-up data.

The increased tunneling currents in these devices, compared with the LW10 devices, was expected given the smaller bandgap. H1RG-18509 was designed to address this concern. A substantial decrease in band-to-band tunneling current in this array compared with the other three arrays can be seen in Fig. \ref{fig:18509_Imodel}. The small curvature in the SUTR curves (shown in the double derivative at the beginning of the SUTR curves in Fig. \ref{fig:curvature_350mV_28K}) for the majority of the pixels for this array is due to the relatively small tunneling current found in the I-V data (Fig. \ref{fig:18509_Imodel} (b)). As a consequence, we are able to achieve larger well depths by applying a large bias ($\sim$75 $ke^-$ with an applied bias of 350 mV). The warm-up data for this array were taken with an applied bias of 250 mV. Dark current data for this array above 29 K is dominated by G-R and diffusion, while lower temperature dark current which is well below 1 $e^-/s$ approaches the light leak level.

The apparent dominance of band-to-band tunneling at higher biases in operable pixels for all arrays is very encouraging, as further enhancements in Teledyne's design to increase $\beta$ will decrease this tunneling component in future longer wavelength devices.

\section{Discussion on Linear Dark Current Regimes}
\label{sect:Discussion}
We have shown that tunneling dark currents dominate the operabiilty of these devices, where the non-linear effects of these currents may present a problem in the calibration of low signal data.

The reduction of tunneling dark currents in H1RG-18509 allows for a linear dark current calibration at low signals since the dark current \textit{vs.} time is nearly linear at biases as large as 350 mV. The non-linear effects on collected signal \textit{vs.} time due to large band-to-band tunneling currents seen in the other three arrays at large biases may be modeled for each of the operable pixels in an attempt to calibrate it, but the exact (or very close to) initial actual bias across each pixel is needed since the tunneling currents vary appreciably as pixels debias.

Alternatively, the three arrays that show large band-to-band currents can be operated in regimes where the dark current is not dominated by band-to-band tunneling.

In the low applied bias regime where band-to-band tunneling is negligible, thermal dark currents would be the dominant source of dark current and could be calibrated at stable temperatures since these currents are approximately linear (above $\sim$25 mV of reverse bias) as pixels debias. For applications that require larger well depths than can be attained with the Hawaii-XRG multiplexers at the biases needed to operate in this regime ($\sim <$ 200 mV), a capacitive transimpedance amplifier (CTIA) multiplexer could be used in its place. A CTIA multiplexer would allow the operation of these arrays at a constant voltage while signal is integrated, therefore maintaining a constant dark current, and with a much larger well depth at the expense of higher power dissipation and read noise.

Another solution to avoid non-linear dark currents as a function of bias is to increase the operating temperature. Tunneling currents decrease with increasing temperature since the bandgap energy increases. At higher temperatures, the dark current due to thermal currents may be comparable to those from band-to-band tunneling at lower temperatures. Figure \ref{fig:18508_Imodel}(a) shows that with an applied bias of 250 mV, the dark current in H1RG-18508 at 35 K is dominated by G-R and is comparable to that at 28 K which is dominated by band-to-band tunneling.

\section{Summary}
\label{sect:summary}
We demonstrate the promising performance of four 13 $\mu m$ cut-off wavelength arrays which all have high quantum efficiency, low read noise, and operabilities close to or greater than 90\% for temperatures of up to 32 K for an applied bias of 250 mV. At lower temperatures the majority of pixels have dark currents below 1 $e^-/s$ for three of the four tested arrays, while the fourth has currents of $\sim$ 10 $e^-/s$ due to a probable glow from the mux.

The cross-hatching pattern associated with the intersection of slip planes and the growth plane in HgCdTe detector arrays was observed for all four LW13 arrays in pixels with high dark currents and/or low well depths due to trap-to-band tunneling dark current. Cross-hatch patterns are generally related to the formation of misfits, generated either during growth or fabrication; further optimization of the process could potentially result in further improvements. Pixels which were operable at a temperature of 28 K and 250 mV of applied bias ($\sim$ 90\% for all four arrays) appear to be dominated by band-to-band tunneling dark current at biases greater than about 200 mV for three of the four arrays. Band-to-band tunneling is also responsible for the inoperability of these three arrays with an applied bias of 350 mV. H1RG-18509 was shown to have an improved performance at larger applied biases with the mitigation of the band-to-band tunneling presumably because of Teledyne's experimental structure, and better operability at lower bias, due to less trap-to-band tunneling. At a temperature of 28 K and an applied reverse bias of 350 mV, the median dark current and well depth for this array is 1.8 $e^-/s$ and 81 $ke^-$ respectively.

These long-wave arrays which can be passively cooled in space will provide cost-savings, offer a longer operational time-frame, and entail no decrease in sensitivity-thereby improving remote sensing capabilities and performance for future space missions. 

TIS has completed the growth of several 15 $\mu m$ cutoff wavelength arrays, and results on those arrays will be reported separately once characterization tests and analysis are complete.

\acknowledgments
The University of Rochester group acknowledge support by NASA grant NNX14AD32G S07. M. Cabrera acknowledges the NASA grant, New York Space grant, and the Graduate Assistance in Areas of National Need (GAANN) grant for partially supporting his graduate work.


\bibliography{LW13_refs}   
\bibliographystyle{spiejour}   

\end{spacing}
\end{document}